%% file: main.tex
\documentclass[a4paper,fleqn,usenatbib]{mnras}
\usepackage{docswitch}
\usepackage{lsstdesc_macros}
\usepackage{tikz}
\usetikzlibrary{calc}
\usepackage{mwe}
\usepackage[htt]{hyphenat}
\usepackage{graphicx}
\graphicspath{{./}{./figures/}}
\extrafloats{100}
\usepackage{lineno}
\usepackage{amsfonts,amssymb,amsmath}
\usepackage{eso-pic}

\AddToShipoutPictureBG*{%
  \AtPageUpperLeft{%
    \hspace{0.75\paperwidth}%
    \raisebox{-4.5\baselineskip}{%
      \makebox[0pt][l]{\textnormal{FERMILAB-PUB-19-618-AE}}
}}}%

\begin{document}

\title[LSST DESC DC1]{The LSST DESC Data Challenge 1: Generation and Analysis of Synthetic Images for Next Generation Surveys }
\input{authors.tex}
\date{\today}
\maketitle
\begin{abstract}
  Data Challenge 1 (DC1) is the first synthetic
  dataset produced by the Rubin Observatory Legacy Survey of Space and Time (LSST)
  Dark Energy Science Collaboration (DESC). DC1 is designed to
  develop and validate data reduction and analysis
  and to study the impact of systematic effects that
  will affect the LSST dataset. DC1 is comprised of $r$-band
  observations of 40 deg$^{2}$ to 10-year LSST depth. We present
  each stage of the simulation and analysis process: a) generation, by synthesizing sources from cosmological
  N-body simulations in individual sensor-visit images with different observing conditions; b)
  reduction using a development version of the
  LSST Science Pipelines; and c) matching to the input cosmological catalog for validation and testing. We
 verify that testable LSST requirements pass within the fidelity of DC1. We establish a
 selection procedure that produces a sufficiently clean extragalactic sample for clustering analyses and
  we discuss residual sample contamination, including contributions from 
  inefficiency in star-galaxy separation and imperfect deblending. 
  We compute the galaxy power spectrum on the simulated field 
  and conclude that: i) survey properties have an impact of 50\% of the statistical uncertainty for the scales and models
  used in DC1 ii) a selection to eliminate artifacts in the catalogs is necessary to avoid biases in the measured clustering;
   iii) the presence of bright objects has a significant
  impact (2- to 6-$\sigma$) in the estimated power spectra at small scales ($\ell > 1200$),
  highlighting the impact of blending in studies at small angular scales in LSST; 


\end{abstract}


\section{Introduction}
\label{sec:intro}
The increase in statistical power of recent cosmological experiments makes the modeling and mitigation of systematic uncertainties key to extracting the maximum amount of information from these surveys. End-to-end simulations~\citep{Brun:118715, AGOSTINELLI2003250, 2006JHEP...05..026S} provide a unique framework to
model systematics and streamline processing and analysis pipelines since they provide a complete understanding of the inputs and outputs. With the increasing availability of computational resources, end-to-end simulations have started to become more prevalent in imaging surveys~\citep{2016ApJ...817...25B, 2018MNRAS.481.1149Z}, and similar efforts are being undertaken in spectroscopic surveys such as the Dark Energy Spectroscopic Instrument~\citep{2016arXiv161100036D}.

One the science goals of the Rubin Observatory Legacy Survey of Space and Time (LSST) is the study of dark energy~\citep{Overview} using probes such as weak lensing (WL), galaxy clustering, clusters of galaxies, supernovae and strong lensing. The increased sensitivity of LSST relative to precursor Stage III dark energy experiments~\citep{2006astro.ph..9591A} necessitates more stringent control over systematic uncertainties for probes like galaxy clustering and WL. The development of end-to-end simulations enables validation and verification of the processing and analysis pipelines. For example, image simulations can be used to evaluate the performance of different shape-measurement algorithms, deblending algorithms, and other processing algorithms. In addition, the expected data volume of LSST, $\sim 50$ PB of raw data and $\sim 40$ billion objects~\citep{Overview} after 10 years, motivates the use of simulated data sets for the development of data handling and analysis pipelines.  

The Dark Energy Science Collaboration
(DESC\footnote{\url{http://lsstdesc.org/}.}) has planned a series of
Data Challenges (DCs) carried out over a period of years, aimed at
successively more stringent and comprehensive tests of analysis
pipelines, to ensure adequate control of systematic uncertainties for
analysis of the LSST data.  An additional goal of these DCs is
development of the infrastructure for analyzing, storing, and serving
substantial data volumes; even while using the outputs of the LSST
Science Pipelines as inputs into analysis pipelines, the
analysis pipelines will need to handle quantities of data greater than that of 
ongoing surveys, even after a single year of the LSST. 
Moreover, it is anticipated that non-negligible subsets of
the data may need to be reprocessed to generate systematic error
budgets (e.g., assessing sensitivity of the results to certain stages
of the analysis process by changing some parameters in the
analysis). These goals will be achieved in practice by a combination
of reprocessing of precursor datasets, which have the advantage of
being fully realistic, and completely synthetic datasets, that have both a known ground 
truth and the ability to enable and disable various effects and thus study them in a more controlled environment.

The goals of the DCs dictate a gradual increase in the
sophistication and volume of the simulated data. In this paper, we
present and analyze simulated images from DC1, the first such
data challenge planned within DESC. The nominal goal 
is to produce synthetic data corresponding to 10 years of integration
in the $r$-band over a contiguous patch of the sky covering approximately 40 deg$^{2}$. 
Only one of the six LSST filters is represented, and the area is 
a small fraction ($\sim$ 0.2\%) of the total LSST area. We
describe how the simulation was achieved and characterize the resulting
products. We validate the basic photometric and astrometric
calibration of these products and check the performance of the
pipeline against the requirements set by LSST and DESC in their
respective Science Requirements Documents~\citep{LPM-17,
  2018arXiv180901669T}. To check suitability of this dataset for galaxy clustering
measurements, we perform a two-point clustering analysis in harmonic space, 
assess the impacts of observing conditions, foregrounds, and detector characteristics as potential sources 
of systematic effects and how the observing strategy can mitigate their impact. Additionally, we assess the importance of sample
selection and how the presence of artifacts and the presence of bright stars can bias the clustering results.
The DC1 data products encompass single-visit and coadded calibrated exposures (i.e., flattened, background subtracted,
etc.) and source catalogs that, together, have a volume of $\sim 225$ TB.

This paper is structured as follows: In \secref{design} we summarize the factors that informed
the design of this data challenge, the inputs, and observing strategies used for DC1. In \secref{image_generation_pipeline} we briefly summarize the tools used to generate and process synthetic images for DC1. \secref{matching} describes the matching procedure to relate the simulation inputs with their corresponding outputs. In \secref{catalogs}, we illustrate the data products generated and perform several validation tests. In \secref{data_selection} we summarize the procedure to obtain a clean sample of galaxies suitable for clustering analyses. In \secref{results}, we present the clustering analyses on the simulated data products. Finally, in \secref{conclusions}, we offer concluding remarks.

\section{Data Challenge Design}
\label{sec:design}

As mentioned in \secref{intro}, the design of DC1 is driven by a
combination of several needs: developing infrastructure for processing and serving data in a way that is useful to DESC and building and testing analysis pipelines including different strategies to mitigate systematic uncertainties affecting various dark energy probes. The philosophy behind the design of the DESC data challenges is to increase the complexity and level
of realism of the datasets in each subsequent iteration. Thus, DC1 is limited in scope and focus, testing a subset of the systematics affecting the different probes that DESC will use.


DC1 serves as a stepping stone to eventually produce simulated surveys covering hundreds to thousands of square degrees in DC2 and beyond.  The area of DC1 is sufficient to enable tests of
two-point clustering statistics up to $\sim 1$ degree scales.  To ensure the volume of simulated images is
tractable, DC1 only includes images in a single band ($r$-band), but goes to full LSST 10-year
depth. 

\begin{figure}
\centering
 \includegraphics[trim={0cm 1.5cm 0cm 1.05cm}, clip, width=1.0\columnwidth]{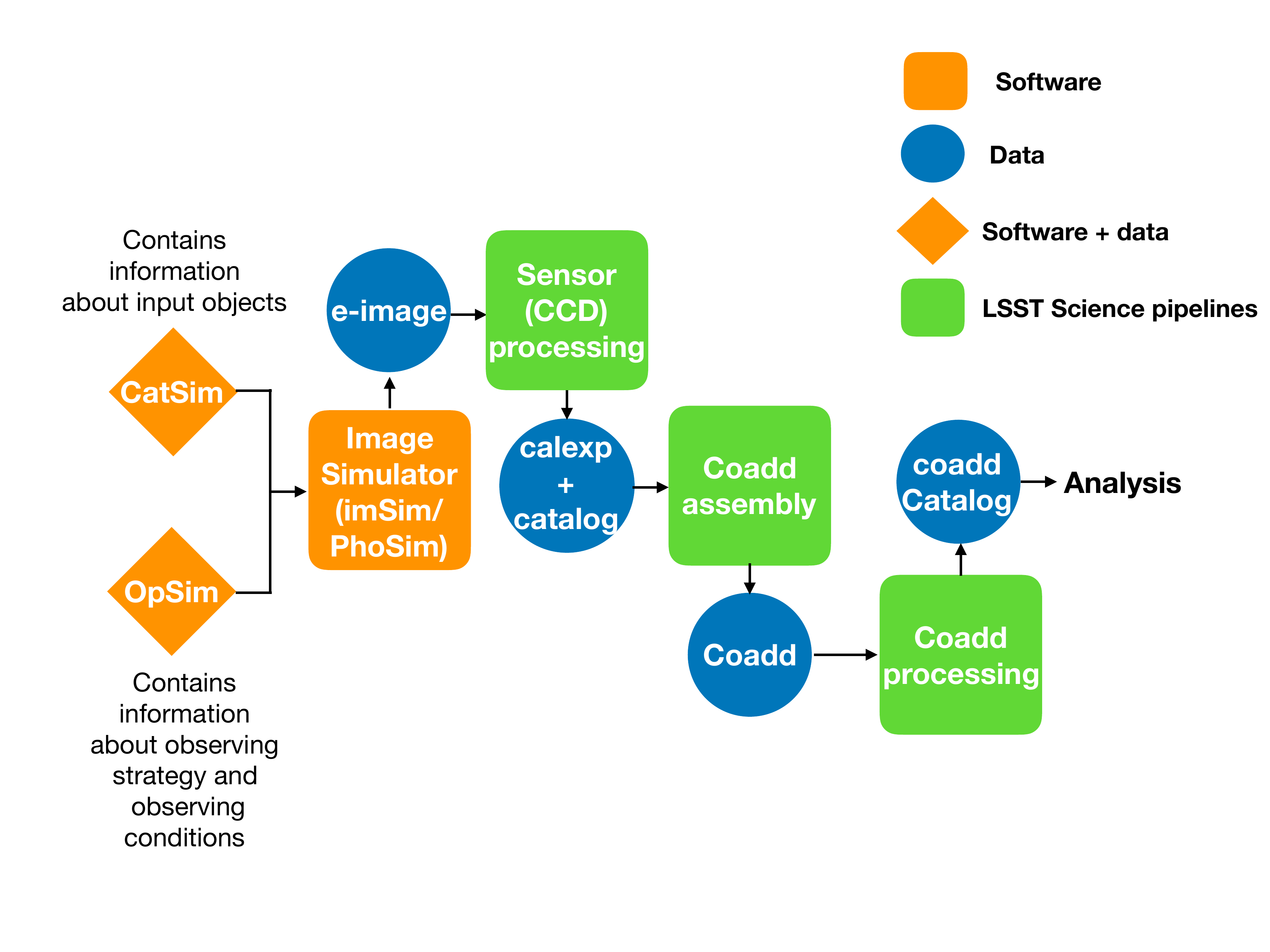}
\caption{Workflow diagram for full data simulations. The image simulation step produces synthetic raw observations from a known truth catalog based on N-body simulations and simulated observing conditions. These data are processed by the LSST Science Pipelines to first generate calibrated single exposure images. These are calibrated both astrometrically and photometrically and are fed again into the LSST Science Pipelines that produce the image co-adds and a catalog of detected static sources. 
}

\label{fig:dc1_workflow}
\end{figure}

The full simulation workflow is depicted in \figref{dc1_workflow}. Briefly, we use as inputs the positions, shapes and fluxes from a galaxy mock catalog from CatSim~\citep{2010SPIE.7738E..1OC,2014SPIE.9150E..14C}, which we describe in more detail in \secref{inputs}, and the observing conditions and strategies described in \secref{dithering} using LSST Operations Simulator~\citep[OpSim][]{2014SPIE.9150E..15D}. These are passed to our image simulation packages described in \secref{imsim_pipeline} that produce raw e-images (i.e., full sensor images in electrons per pixel without any added instrumental effects such as cross-talk, bleeding, etc.). These e-images are then processed by the LSST Science Pipelines~\citep{2015arXiv151207914J, 2018PASJ...70S...5B}. The processing is described in \secref{image_processing_pipeline}. These produce the calibrated exposures, coadds and catalogs that we use for our analysis.

\subsection{Image generation: input catalog}
\label{sec:inputs}
\begin{figure}
\centering
\includegraphics[width=0.9\columnwidth]{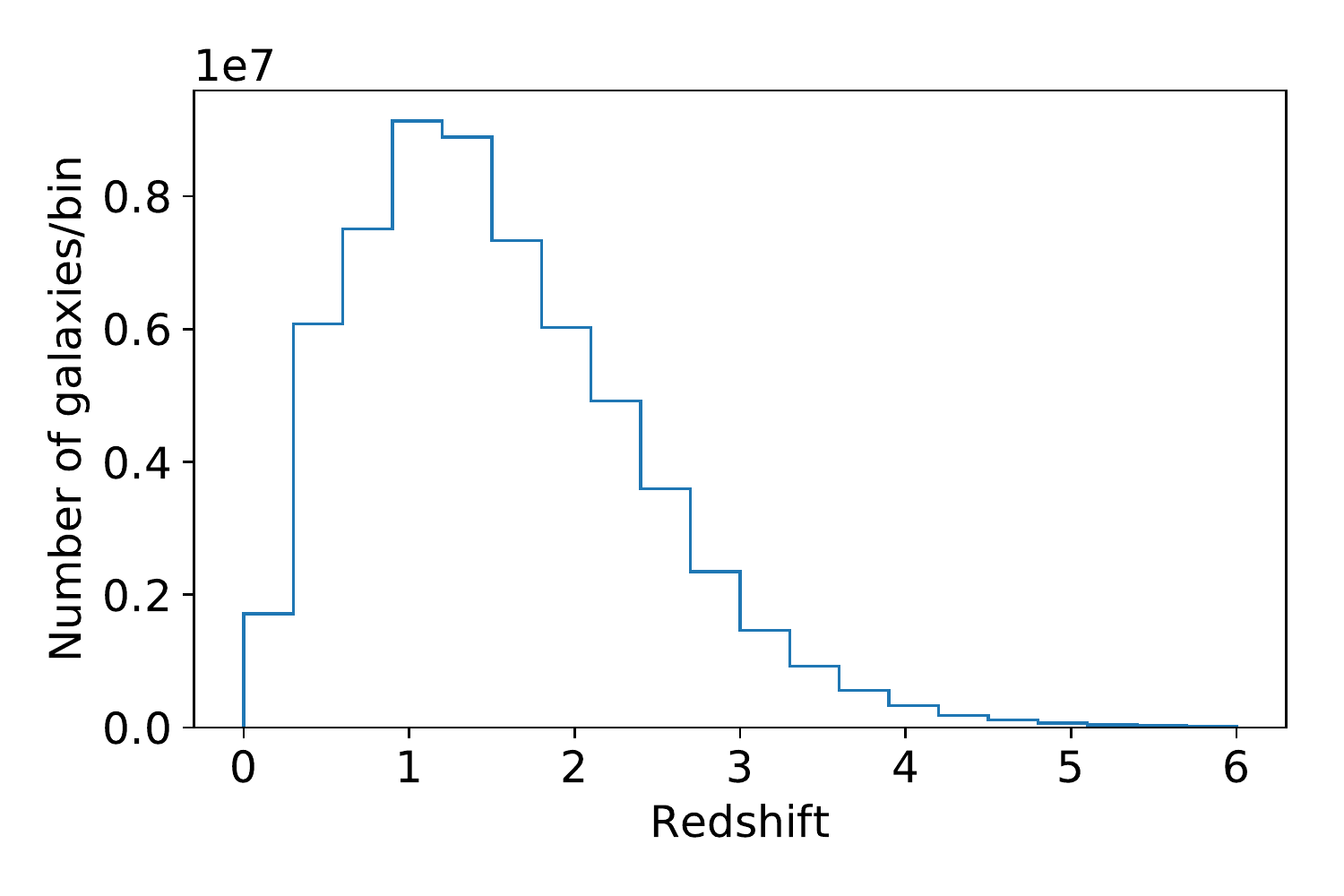}
\includegraphics[width=0.9\columnwidth]{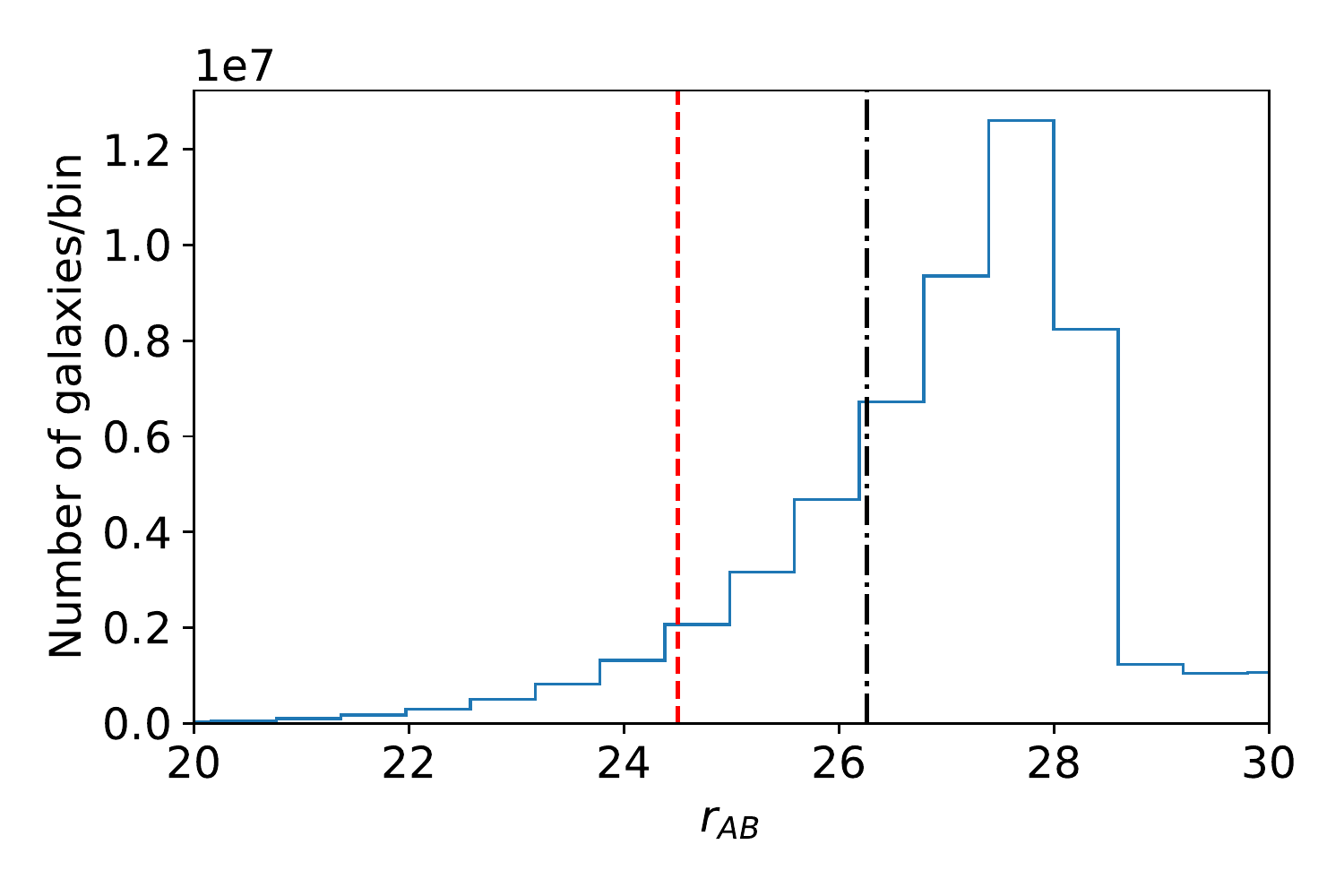}
\caption{Redshift (top) and magnitude (bottom) distribution for the galaxies used as inputs for the Data Challenge 1 simulations. In the magnitude distribution we include, as references, the typical 5-$\sigma$ point source detection depth for a single exposure (red dashed line) and the median 5-$\sigma$ point source detection depth in the deepest coadded DC1 simulation (black dashed-dotted line).}
\label{fig:catalog_plots}
\end{figure}

Image simulations allow us to assess the detection and deblending performance of a given image-processing pipeline. For example, if we produce images using an object catalog with random positions uniformly distributed across the sky, as well as uniformly random shapes and fluxes, we can get information about detection efficiencies as a function of flux. However, the effects of source blending would not be realistic as we would not be able to capture some correlations present in real data. On the other hand, using N-body simulations as the input to generate artificial images allows us to study all the aforementioned effects. We used the CatSim~\citep{2010SPIE.7738E..1OC,2014SPIE.9150E..14C} catalog as our input to include realistic correlations between galaxies and to be able to test our analysis pipelines. CatSim is a set of catalog-level simulations provided by the LSST Simulations Team representing a realistic distribution of both Milky Way and extragalactic sources. In particular, the extragalactic catalog contains galaxies spanning the redshift range $0 < z < 6$ in a 4.5 deg$\times$4.5 deg footprint. The magnitude and redshift distributions are shown in \figref{catalog_plots}. The galaxies in CatSim are generated by populating the dark matter haloes from the Millennium simulation~\citep{2005Nature.435.629S} using a semi-analytic baryon model described in \citet{2006MNRAS.366..499D} including magnitudes BVRIK, LSST-ugrizy, and bulge-to-disk ratios. For all sources, a spectral energy distribution (SED) is fit to the galaxy colors using \citet{2003MNRAS.344.1000B} spectral synthesis models. Fits are undertaken independently for the bulge and disk and include inclination-dependent reddening. Morphologies are modeled using two S\'{e}rsic profiles~\citep{1963BAAA....6...41S} and a single point source (for the AGN). Half-light radii for the bulge components are derived from the absolute-magnitude vs. half-light radius relation given by \citet{2011A&A...534A...3G}. Stars are represented as point sources and are drawn from the Galfast model~\citep{2008ApJ...673..864J}. More information about these catalogs can be found at the LSST Simulations webpage\footnote{\url{https://www.lsst.org/scientists/simulations/catsim}.}. In principle, the semi-analytic model used in CatSim can give us insight about the small-scale regime that LSST may be sensitive to. However, at fainter magnitudes $(r > 25)$ the semi-analytic model underpredicted the overall number density of objects, and unclustered galaxies were added in order to obtain the correct number densities (Connolly, private comm.) which can potentially dilute the signal. We do not expect that using a different galaxy inpainting model would change the conclusions of this work since we do not attempt to make any measurement of cosmological or halo-occupation-distribution parameters.

For DC1, we chose a nominal field centered at RA $\approx 93^{\circ}$ and Dec $\approx -29^{\circ}$. This field has a Galactic latitude of $b \approx -20^{\circ}$ and a dust extinction per magnitude of interstellar reddening $0.05 \leq E(B-V) \leq 0.35$ and thus represents a typical region in the LSST wide-fast-deep survey~\citep{Overview}. The CatSim catalog was tiled to generate a 40 deg$^{2}$ footprint, covering 4 LSST full focal plane pointings. This approach introduces a periodicity that induces extra correlations in our sample; however, this is not a major issue as with the relatively small area of the DC1 simulation we are unable to measure correlations on large scales ($> 1$ deg) with any useful precision.


After tiling, the input catalog contains approximately $63.1$ million sources, of which 97\% are galaxies whose redshift and magnitude distributions are depicted in \figref{catalog_plots} and the remaining objects are stars. We simulate $r$-band observations to the LSST full depth ($10$ years, 30-second exposures) within the DC1 footprint using an observing cadence generated with OpSim\footnote{Specifically, we use the \texttt{minion\_1016} database: \url{https://www.lsst.org/scientists/simulations/opsim/opsim-v335-benchmark-surveys}.}~ \citep{2014SPIE.9150E..15D}, which contains simulated pointing position, observation date and filter. It also contains information about simulated observing conditions, such as seeing, sky-brightness, and moon position. For DC1 we use 4 pointings with $\approx 184$ visits per pointing over 10 years from this database.

\subsection{Dither strategy}
\label{sec:dithering}

OpSim's output contains a realization of the LSST observing cadence and the survey footprint. Since OpSim divides the sky into hexagonal tiles, the nominal telescope pointings lead to overlapping regions across adjacent tiles that are observed more often than the non-overlapping part of the field of view (FOV), resulting in depth non-uniformity on the scale of $\sim$1 degree. This non-uniformity can introduce systematic uncertainties in the two-point statistics of galaxies~\citep{2016ApJ...829...50A}. In an effort to mitigate these effects, and following the same approach that will be taken with LSST data, we implement \textit{dithers} -- offsets in the nominal telescope pointings. Specifically, here we use \textit{large}, i.e., as large as half the FOV, random translational and rotational dithers implemented for every visit. Note that these dithers differ from those recently discussed in~\citet{2018arXiv181200515L}. This translational dither strategy is chosen based on a more extensive study of the various (translational) dither strategies in \citet{2016ApJ...829...50A}, where random dithers for every visit are found to be among the most effective.

We consider both an undithered and a dithered observing strategy. For the dithered strategy, some visits contain sensors that fall outside of the DC1 region; these sensors were not simulated in order to save computational resources. However, the sensors that partially overlapped our nominal field of view were simulated. In total, we simulate $\approx 184,600$ sensor-visits for the dithered simulation and $\approx 151,000$ for the undithered simulation.

\section{Image generation and processing}
\label{sec:image_generation_pipeline}

The artificial generation of astronomical images is a complex and computationally demanding process. In recent
years, there have been major efforts in the community to create software that enables the generation of astronomical images, with various choices made in terms of level of complexity, fidelity, and computational efficiency, such as PhoSim\footnote{\url{https://bitbucket.org/phosim/phosim_release/wiki/Home}.}~\citep{2015ApJS..218...14P}, and \texttt{UFIG}\footnote{\url{https://cosmology.ethz.ch/research/software-lab/ufig.html}.}~\citep{2016ApJ...817...25B}. For DC1, we model the input sources using
two independent approaches. 

Firstly, we use PhoSim, a fast photon Monte Carlo code that enables the generation of images with a high level of realism and can use LSST-specific information (e.g., the geometry of the CCDs and the focal plane, the system throughputs in the different bands, etc.) to generate LSST-like images. We use PhoSim version 3.6 with a custom configuration that enables some
approximations (photon bundling) to be made when generating the sky background in order to reduce the overall computing time needed to produce the images.

We also employ imSim\footnote{\url{https://github.com/LSSTDESC/imSim}.} (Walter et al., {\em in prep.}), which internally uses GalSim~\citep{2015A&C....10..121R} as a library for image rendering, and also uses LSST-specific information to generate synthetic images. For DC1 we use an early pre-release version of imSim, version \texttt{v.0.1.0}, which performs a series of approximations that allow us to complete the image simulations significantly faster ($\sim 60\times$) than PhoSim v3.6, at the expense of a loss in realism. These approximations include simplifications in the point-spread function (PSF) model (PhoSim performs ray-tracing through the atmosphere, while this version of imSim uses simple parametric models for the PSF) and omission of all sensor effects, which were deemed acceptable given the goals of DC1.

Due to the computational resources needed to run PhoSim we could only generate one campaign of the dithered DC1-PhoSim data, whereas we could produce the dithered and undithered campaigns for imSim. Furthermore, comparison of the results for the PhoSim and imSim images has proven less informative than intended due to some unusual features of unknown origin in the sky backgrounds in the PhoSim images\footnote{Several aspects of the implementation of the sky background rendering -- including the photon-bundling approximations -- are updated in subsequent versions of PhoSim, so this finding may not carry over to later PhoSim versions.}. For these reasons, we focus on the analysis and comparison of dithered versus undithered imSim images for the rest of this work.

\subsection{DC1 imSim configuration}
\label{sec:imsim_pipeline}


For DC1, we use imSim to simulate each CCD of the focal plane individually, and generate a single image with a 30-second exposure time. We omit sensor effects and variability in the optical model across the focal plane. We set the gain to be 1. Our sky brightness model is based on the \citet{1991PASP..103.1033K} model provided by OpSim, refined with the detailed wavelength dependence of the phenomenological model from~\citet{2016SPIE.9910E..1AY}. The PSF model is a Gaussian for the system (telescope, CCD and other elements that may be in the optical path other than the atmosphere) with a  dependence on airmass, $X$, of the full-width half-maximum, $\mathrm{FWHM_{sys}} = (0.4\arcsec) X^{0.6}$, to approximate the degradation in the image quality due to, e.g., gravity loading\footnote{See LSE-30~\url{http://ls.st/lse-30} p.~80.}. An airmass-dependent Kolmogorov profile\footnote{\url{http://galsim-developers.github.io/GalSim/_build/html/_modules/galsim/kolmogorov.html}.} is used to model the atmosphere. In order to be consistent with the sky brightness model we use an airmass, $X$, model that depends on the angular distance to the zenith, $Z$, from~\citet{1991PASP..103.1033K}:

\begin{equation}
X = (1 - 0.96\sin^{2}{Z})^{-0.5}.
\end{equation}

For DC1, imSim is used to generate three different types of objects: stars, which are modeled as PSF-like objects; galaxies, which are modeled as composite (bulge plus disk) S\'{e}rsic profiles~\citep{1963BAAA....6...41S} using
the parameters given by CatSim; and AGNs which are also modeled as point sources and, for simplicity, without any variability.\footnote{Newer versions of imSim have the ability to generate more complex galaxy morphologies (e.g., they can include random Gaussian knots).} The brightness for these sources is computed using the magnitudes from CatSim, which are converted to counts using the latest version of the LSST throughputs\footnote{\url{https://github.com/lsst/throughputs}.}. In DC1, we clip the objects at magnitude 10 in order to improve the computational efficiency. Saturation is emulated by clipping the maximum number electrons per pixel in the CCDs at 100,000.

The final products of the image generation process are 4k $\times$ 4k pixel images (in FITS format). We generated more than 300,000 sensor-visit images in total (including both the \textit{dithered}, and \textit{undithered} fields). The average time to simulate each CCD is $\sim 4000$ seconds and the total production time is $\sim f0,000$ CPU-hours.

\subsection{Image processing}
\label{sec:image_processing_pipeline}

\begin{figure*}
\centering
\includegraphics[width=0.45\textwidth]{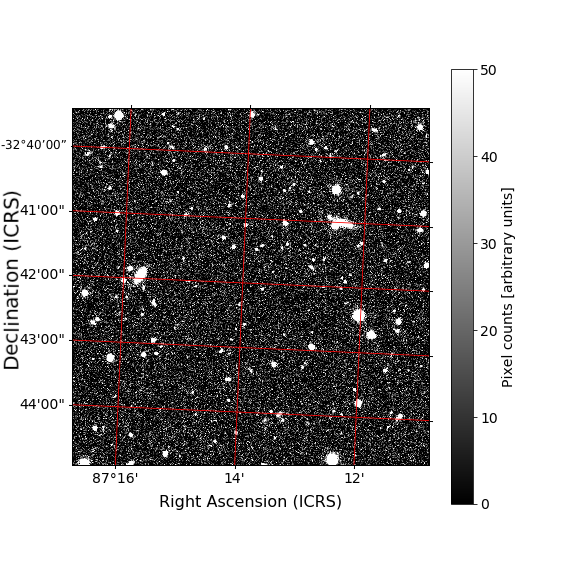}
\includegraphics[width=0.45\textwidth]{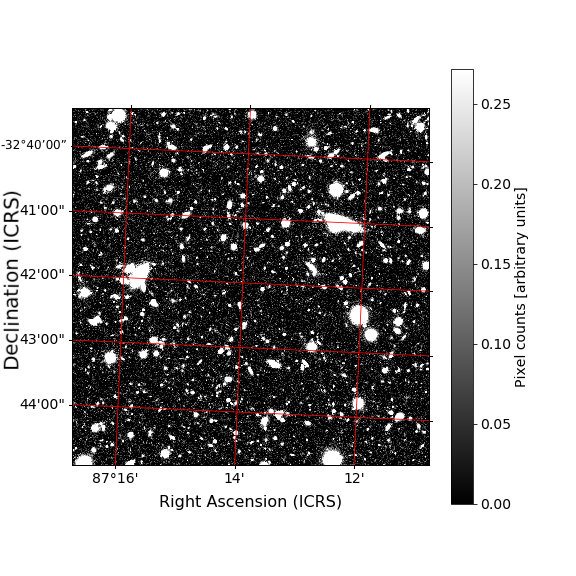}
\includegraphics[width=0.45\textwidth]{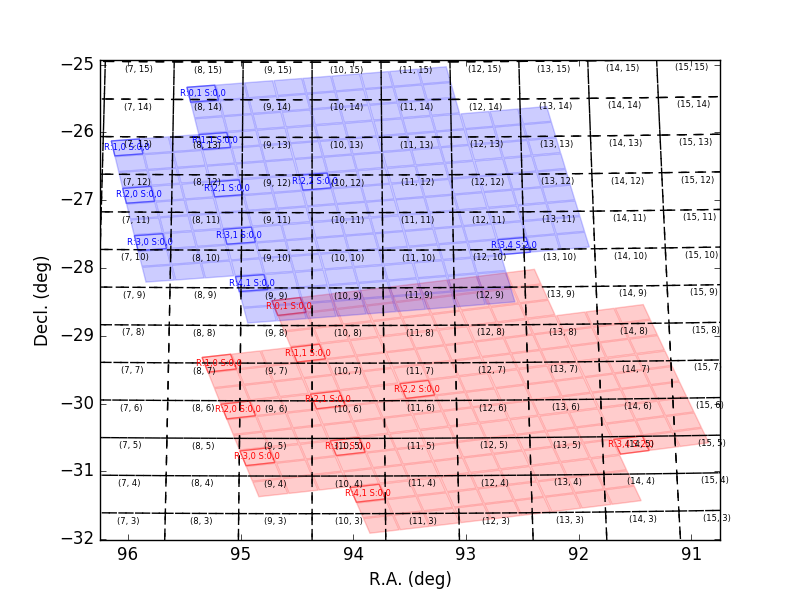}
\includegraphics[width=0.45\textwidth]{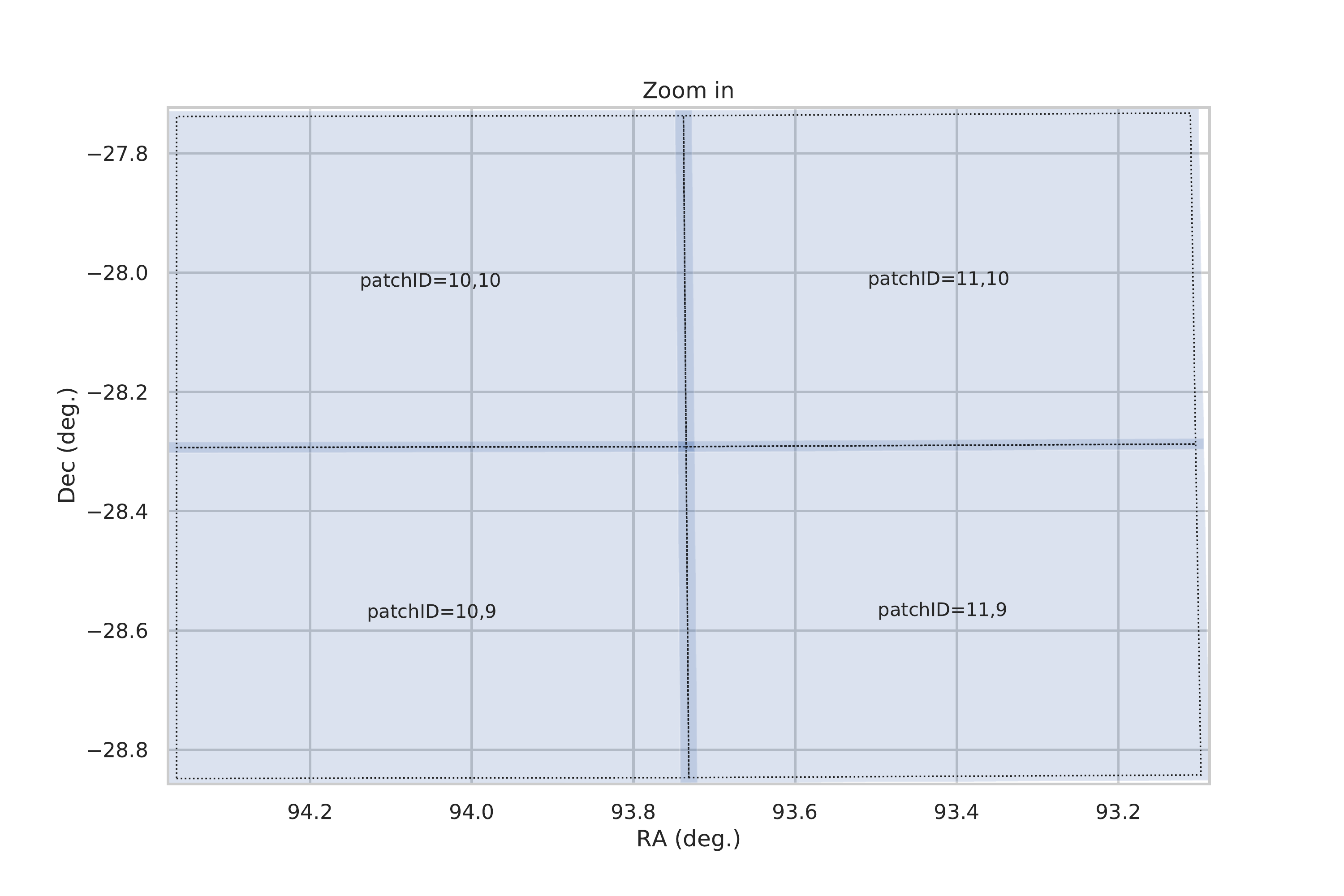}
\caption{Example of a 1000 $\times$ 1000 pixel cutout from a calibrated exposure (top left), i.e. background subtracted, reduced single-epoch image; a full depth coadd (top right). We can see the stark difference in the number of objects that are detectable by eye. The red grid corresponds to lines of constant Right Ascension (vertical lines) and constant declination (horizontal lines). We also show a visualization of the coaddition scheme used by the Science Pipelines (bottom left and right). In the bottom left panel the different images in each sensor visit (red and blue squares) are mapped to patches (blank squares delimited by dashed lines) where the coaddition is performed independently. These patches have small 1 arcmin-wide overlaps with each other, and each patch is identified by a pair of numbers. The bottom right panel shows nine different patches as light blue squares, and we can see their overlap as the darker blue areas.}
\label{fig:coadd_example}
\end{figure*}

The outputs of these simulations are processed using the LSST Science Pipelines~\citep{Overview,ScienceBook,WhitePaper,2015arXiv151207914J,2018PASJ...70S...5B} version 13.0\footnote{\url{https://pipelines.lsst.io/releases/v13_0.html}.}, which we will refer to as the Data Management (DM) stack. The DM stack is an open-source, high-performance data processing and analysis system intended for use in optical and infrared survey data\footnote{The code can be found at \url{dm.lsst.org} and \url{pipelines.lsst.io}.}. A brief schematic of some of the steps in the image processing pipeline can be seen in \figref{dc1_workflow} as green squares. The raw, uncalibrated single exposures are used as inputs. The software performs the reduction, detection, deblending and measurement on individual visits. It then combines the single-visit images to produce the so-called coadds\footnote{For more information about coaddition, see Section 3.3 in~\citet{2018PASJ...70S...5B}}. This is done by computing a weighted average of resampled overlapping sensor images in a given region of the sky that we call a \texttt{patch}. An illustration of how this process works is shown in~\figref{coadd_example}. After assembling the coadded images, the DM stack performs measurements on them to produce a catalog. The DM stack provides calibrated images and source catalogs for the individual visits and coadds stored in FITS files. In total, we detect and measure $\sim 10.6$ million  (9.7 million for the undithered simulation) objects with position, flux and shape information. We activated optional extensions for the pipeline to include \texttt{CMODEL} fluxes (see \citealt{2018PASJ...70S...5B} for more details) and \texttt{HSM} shapes~\citep{2003MNRAS.343..459H,2005MNRAS.361.1287M}. In \figref{coadd_example} we show an example of a single-visit processed image and a coadded image, which corresponds to 184 dithered overlapping single-visit images. We can see the stark difference in the detectable number of objects just by eye ($\sim 5$ times more objects with SNR$>5$), due to the increased depth (the coadd images are $\sim 3$ magnitudes deeper). 

The reduction pipeline is essentially the same as for Hyper Suprime-Cam (HSC). This allows us to use the HSC selection criteria~\citep[Sec. 5.1]{2018PASJ...70S..25M} as the basis for our analysis, and can potentially enable direct comparisons between datasets for further validation.

The total processing time for the DC1 simulated images is $\approx 29,000$ CPU-hours.

\section{Matching inputs and outputs}
\label{sec:matching}

Using end-to-end simulations, one can potentially trace each measured photon from its corresponding source and fully characterize the image generation and measurement processes. In practice, this is very difficult due to the large data volume and the fact that the data reduction pipeline is built around pixelated images rather than tagged photon counts.

Nevertheless, the output catalog is a noisy representation of the underlying input (truth) catalog and we need to find a way to connect the two. 
The simplest way to associate members of two catalogs is by using the positions of the objects in the sky. This approach has been extensively used in the literature~(e.g., \citealt{1977A&AS...28..211D,1983Obs...103..150B,1986MNRAS.223..279W}) and performs reasonably well when blending is low (i.e, when there are few overlapping sources in the image). However, when the blending fraction is high, this approach might be insufficient. In this case, matching other quantities like flux, color and/or shape can become useful~\citep{2008ApJ...679..301B, doi:10.1146/annurev-statistics-010814-020231}. However, when adding other quantities, the matching process can become slower and result in a lower matching completeness. For more details about challenges relating two different catalogs, we refer the reader to~\citet{doi:10.1146/annurev-statistics-010814-020231}. 

We compare two different matching strategies: positional matching, where for each detected object we find the closest object in the truth catalog, which we will refer to as \textit{pure spatial matching} and will denote as \textsf{S}; and positional matching with magnitude matching, which we will refer to as \textit{spatial+magnitude matching} and denote as \textsf{S+M}, where for each detected object we find objects from the truth catalog that lie within a three pixel radius ($0.6 \arcsec$). After this, we select the object that is closest in magnitude as long as the difference in r-band magnitude, $\Delta r$, is less than a certain threshold. In our case, we conservatively choose $|\Delta r| < 1.0$. Using this approach, if none of the neighbors fulfill these conditions, the detected source is considered unmatched.

For both approaches, we build a \texttt{KDTree}~\citep{scikit-learn} using the positions of detected objects flagged with \texttt{detect\_isPrimary=True} which ensures that the source cannot be further deblended by the pipeline and was detected in the inner part of a patch. The inner part of a patch is the part of the sky exclusive to the said patch. In order to speed up the processing and to reduce the usage of computational resources we build the \texttt{KDTree} using sources from 30 randomly selected patches ($\sim 10\%$ of the total number of patches) in the dithered simulation containing 975,605 detected sources fulfilling the aforementioned condition (we will refer to these as primary outputs or primary detected sources). Using this sample, we find that 95.2\% of the sample is matched using the \textsf{S+M} matching. The undithered simulation yields similar results and conclusions. 

One interesting metric is the fraction of primary detected sources that have been matched to the same object in the input catalog, which we will denote as $f_{\mathrm{multi}}$. This is an unusual occurrence, but a situation where this can happen is when bright objects appear ``shredded", i.e., one bright object is detected as multiple fainter sources. This happens because noise fluctuations of the said bright object are considered as single sources. We find that these kind of matches are $\sim 100$ times more likely to happen using the \textsf{S} matching ($f_{\mathrm{multi}} = 3 \times 10^{-3}$) than in the \textsf{S+M} matching ($ f_{\mathrm{multi}} = 2 \times 10^{-5}$). This is expected because, in the cases where the primary detected source is a random fluctuation or a shredded source, it is unlikely that the measured flux is close to the flux of the true, neighboring source thus producing an unmatched source in the case of using \textsf{S+M} matching. We also find that the two approaches select the same matching source in the truth catalog for only 68\% of the primary detected sources for which we found a match. These differences have several potential explanations: for example, objects with a poorly determined centroid position that are close to other objects in the truth catalog (remember that the truth catalog contains objects up to $r=28$), objects with poorly determined fluxes (low SNR), etc.

\begin{figure}
\centering
\includegraphics[width=0.85\columnwidth]{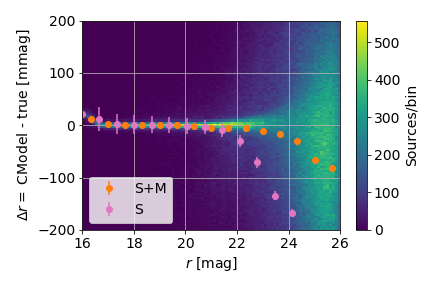}
\caption{Median per-bin photometric residual as a function of measured magnitude for pure spatial matching (pink), and spatial+magnitude matching (orange) using 15 magnitude bins between $r$-band magnitude 16 and 26. We also show the 2D histogram of the residuals for the matched objects using \textsf{S+M} matching as a function of magnitude.}
\label{fig:matching_comparison}
\end{figure}

We compare the photometric residuals, $\Delta r = $ \texttt{CMODEL} - $r_{true}$ (i.e. measured minus input magnitudes), using both approaches in~\figref{matching_comparison}. The resulting median photometric residual seems strongly biased in the case of using \textsf{S} matching. However, in the case of \textsf{S+M} matching, the median residuals and their uncertainties are smaller, as expected given the limit in the magnitude difference. We can see that in this case the biases are still significant, partially because of the fact that faint sources just below the detection threshold are detected only if they have positive noise fluctuations, and we lose some faint sources above the detection threshold that have noise fluctuations that make them appear dimmer, biasing the overall residual distribution. On top of that, spurious matches also contribute to this effect. This can be mitigated using a smaller tolerance in magnitude difference (for example using 0.5 mag instead of 1 mag) but that would result in an overall reduction of the number of matched primary detected sources. Finally, observing the magnitude difference between inputs and outputs of individual matched sources using the \textsf{S+M} technique, we can see that this distribution is centered around zero, except for very bright sources ($r < 17$), where saturation prevents us from accurately determining the fluxes. 

Different matching techniques have different potential applications and strengths~\citep{doi:10.1146/annurev-statistics-010814-020231}. In our case, we want to use these matching techniques to provide a clean (flux-limited) sample to perform two-point clustering analyses. Given that magnitude precision and accuracy will be important for our sample selection, the \textbf{spatial+magnitude} matching technique will be sufficient to clean the sample from spurious and poorly measured sources. However, more complicated matching techniques may be necessary for other use cases.
 
\section{Output catalogs and validation}
\label{sec:catalogs}

In order to check the level of realism and the accuracy of the processed catalogs we perform several validation tests. These tests check two different aspects: the level of realism and consistency of the simulated products with the inputs, and the performance of the simulation and processing pipelines. As a guide to check the status of our end-to-end pipeline, with a special focus on the performance of the processing pipeline, we use the LSST Science Requirements Document\footnote{\url{https://ls.st/LPM-17}.}~\citep{LPM-17}, which we will refer to as the LSST-SRD\footnote{We use the LSST-SRD version 11.}. It describes science-driven requirements for LSST data products in its Key Performance Metrics (KPMs). The results in this section are presented for the dithered simulation; however, unless stated otherwise, the procedures and results of the validation checks are similar for the undithered simulation. 

After performing some basic sanity checks (e.g., of the magnitude distribution, the number density of detected objects, and the footprint coverage) we test our simulations against the different KPMs by processing the output individual-visit and coadd catalogs with the \texttt{validate\_drp}\footnote{This is an open-source package that tests data release products against some KPMs: \url{dmtn-0008.lsst.io}, \url{https://github.com/lsst/validate_drp}.} package. For quick reference, a brief description of the requirements from the LSST-SRD studied in this section can be found in Appendix~\ref{app:lsst_srd}. In addition, we also validate our dataset using some PSF-related requirements in the DESC Science Requirements Document~\citep{2018arXiv180901669T}, which we will refer to as the DESC-SRD.
 
Note that some of the performance requirements for LSST are not met by design in our simulations, e.g., requirements involving colors or more than one filter. We will ignore such requirements in this work. On the other hand, our images automatically meet some criteria due to the design choices (e.g. the pixel size is fixed in our simulations). 

\subsection{Astrometric performance}
\label{sssec:astrometry}

\begin{figure}
\centering
\includegraphics[width=0.9\columnwidth]{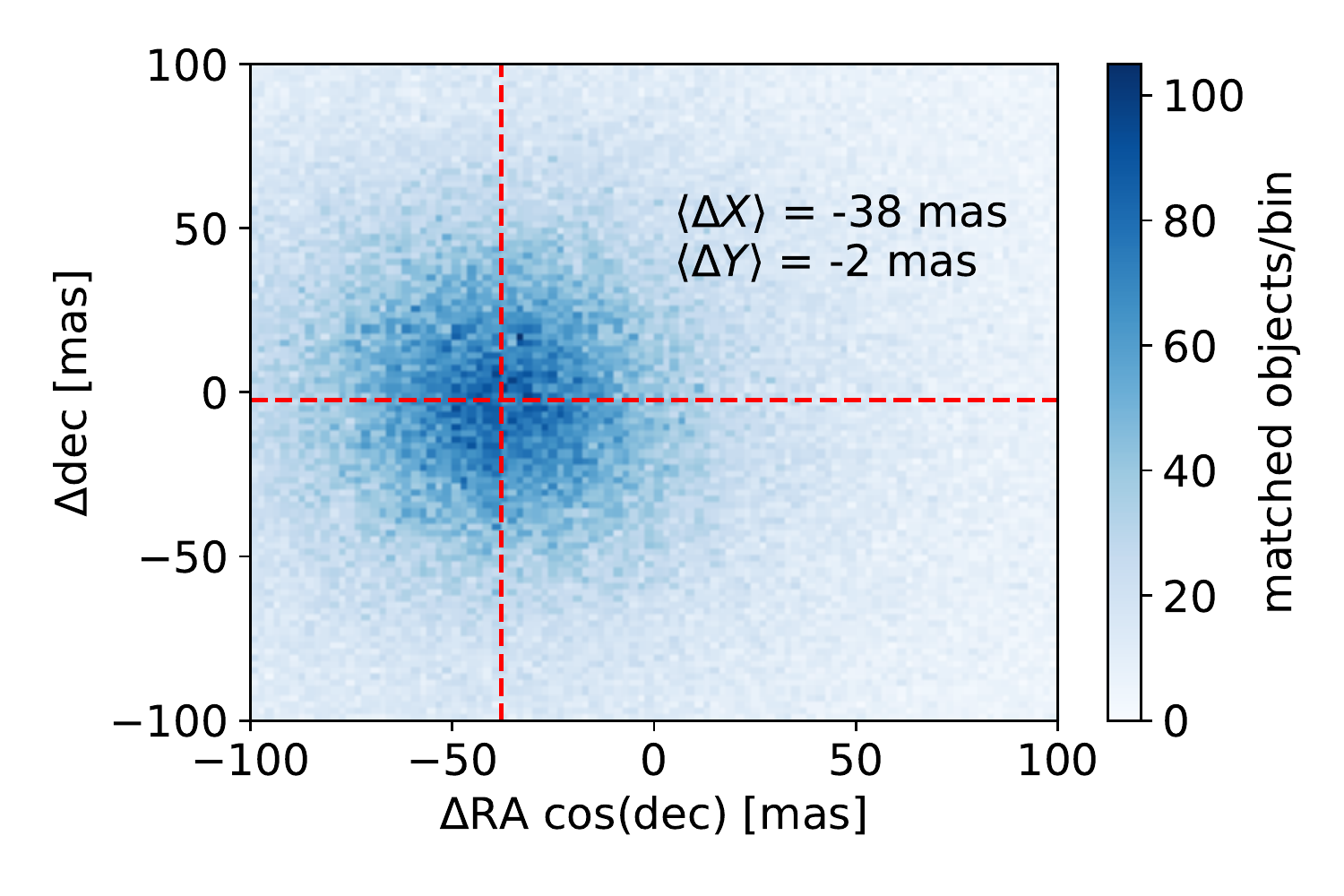}
\caption{Astrometric residuals of detected objects (stars and galaxies) of 500 randomly selected visits in the dithered simulation. We find a bias of 38 mas in the measured right ascension due to uncorrected proper motion. We obtain similar results using the undithered simulation.}
\label{fig:AA1}
\end{figure}

Consistency with a reference frame (absolute astrometry) is required in order to compare with external catalogs. One can think about the case of cross-correlating with an external catalog, e.g. CMB maps; if the astrometric solutions are not consistent the cross-correlation signal will be biased. Much more stringent constraints for absolute astrometric accuracy come from the orbit computations for solar system objects that will be studied with LSST, for which the maximum acceptable mean deviation with respect to the reference is 100 milliarcseconds (mas). We check this requirement by measuring the difference between input and output positions, shown in~\figref{AA1}, finding a median (and mean) deviation of 38 mas. This bias is due to the uncorrected proper motion of stars used for calibration. The proper motions of stars were included in the simulations; however, the catalog used for calibration had the location of these stars fixed at a specific epoch (J2000). This will be solved in future data challenges.

On the other hand, we require that the positions of objects in different visits be consistent (astrometric repeatability). For example, the coaddition process relies on the astrometric solutions for the single-epoch images in order to get deeper images. If the astrometric solutions between different epochs are systematically offset or their errors are too high, we can obtain galaxy profiles that are too wide, or inconsistent relative distances between objects at different epochs, which will affect our clustering and weak lensing science. The study of parallax and proper motion of stars set a more stringent requirement on the astrometric repeatability of LSST. In particular, we check pairs of bright stars $(17.0 < r < 21.5)$ separated by $4.< D < 6$ arcmin, and compare the distribution of the variation in their separation distance at different epochs. We require that the root mean square (RMS) of this distribution should be below 20 mas, and that the fraction of outliers, i.e., pairs where the variation of the separation is larger than 40 mas should be below $20\%$. We repeat this for pairs with $19 < D < 21$ arcmin following the same criteria, and pairs with $199 < D < 201$ arcmin, where we require that the RMS should be below 30 mas, and less than 30\% of the pairs vary their separation distance by more than 50 mas. We find that our dataset passes all of these astrometric requirements, guaranteeing that the positions measured in DC1 will be useful for clustering analyses. 



\subsection{Photometric performance}
\label{sssec:photometry}
\begin{figure}
\centering
\includegraphics[width=0.9\columnwidth]{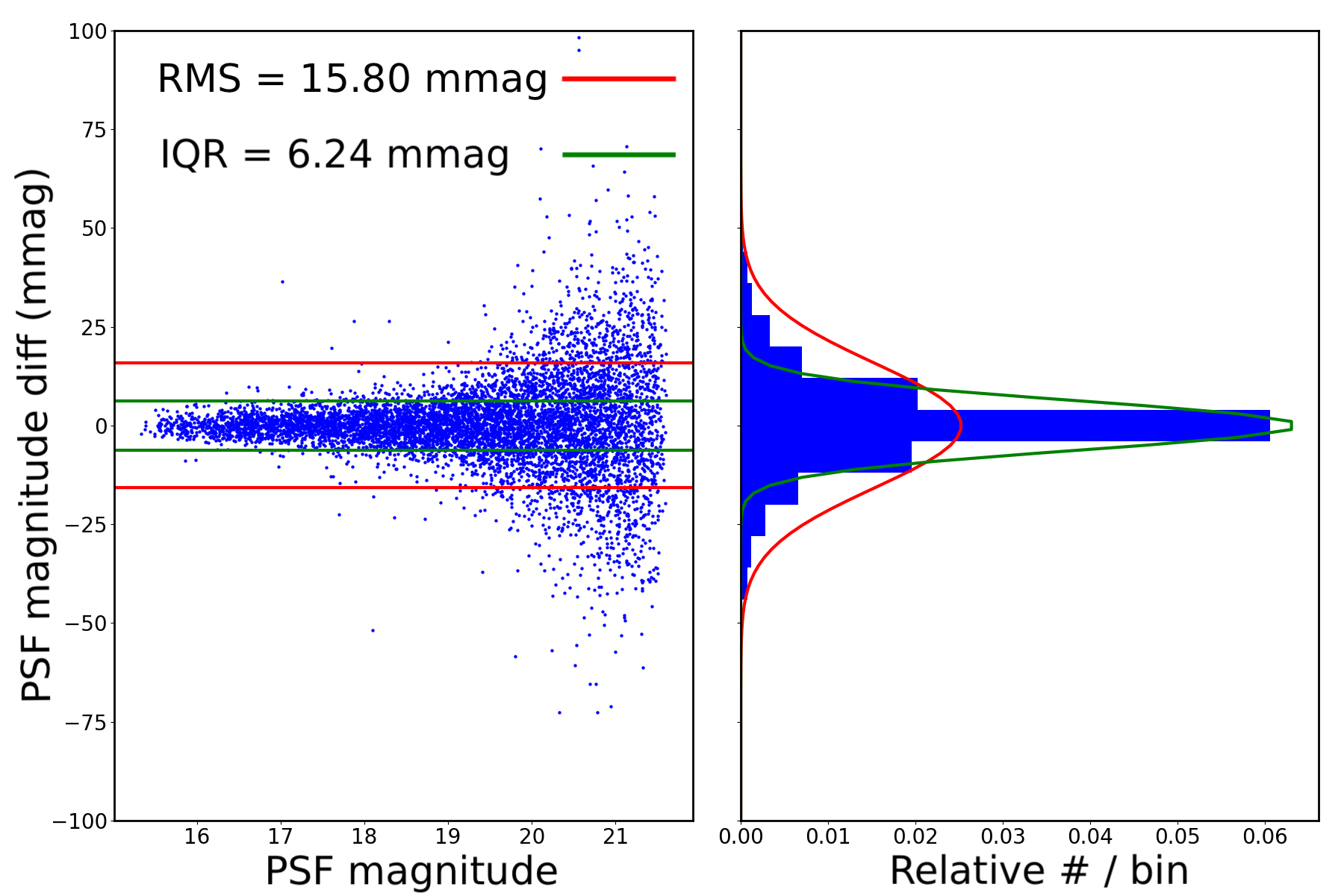}
\caption{(left) The magnitude differences of pairs of measurements of stars across visits for stars with a typical SNR $>100$ as.a function of their measured psf magnitude, i.e., the magnitude measured in a PSF-like aperture.  (right) The histogram of these differences.  The Gaussian root mean square (RMS) is shown in red while the interquartile range is shown in green. Note that the distribution is more peaked than a Gaussian. The interquartile range (6.2~mmag) is {\em smaller} than the Gaussian RMS (15.8~mmag). This means that the distribution has extended tails but most objects have very accurate magnitude measurements.}
\label{fig:validate_drp_PA1}
\end{figure}

In order to have well calibrated supernova lightcurves, as well as accurate photometric redshifts, we need consistent photometric measurements across different visits, since artificial variations due to miscalibration can lead to biased cosmological estimates. In particular, LSST requires that the RMS around the mean magnitude value for bright stars should be below 8~mmag, and that no more than 20\% of bright point-like sources deviate by more than 15~mmag. We check these criteria by comparing the measured magnitudes of bright (SNR $> 100$) point-like objects across different visits as shown in \figref{validate_drp_PA1}. This test validates that the pipeline is reconstructing fluxes of objects consistently across epochs, and also that different epochs are produced consistently in our image simulations. We estimate that the RMS in our case is 15.8~mmag, larger than the requirement; however, we see that the RMS is affected by the presence of outliers and decide to compare with the scaled interquartile range\footnote{we use a rescaled IQR$\equiv$ 75th percentile minus 25th percentile, and then divided by $2\sqrt{2}\mathrm{Erf}^{-1}(1/2)$ (the interquartile range of a Gaussian distribution with $\sigma=1$) in order to have a robust estimation of $\sigma$.}, for which we obtain an $\mathrm{IQR}=6$~mmag; We also check the fraction of photometric, bright point-like sources that deviate by more than 15 mmag, finding only 17\% over this threshold, in compliance with the requirements (20\%). The smaller interquartile range compared to the RMS in combination with the small fraction of outliers suggest that the majority of objects in DC1 have well-measured photometry, while a small fraction of outliers have poorly determined photometry. We do not expect this to cause any biases in clustering or weak lensing measurements. In the case of DC1, these criteria are not crucial since we do not perform any photometric redshift measurements given that we only have single-band imaging.

\subsection{Zeropoint uniformity}
\label{sssec:zeropoints}
\begin{figure}
\centering
\includegraphics[width=0.85\columnwidth]{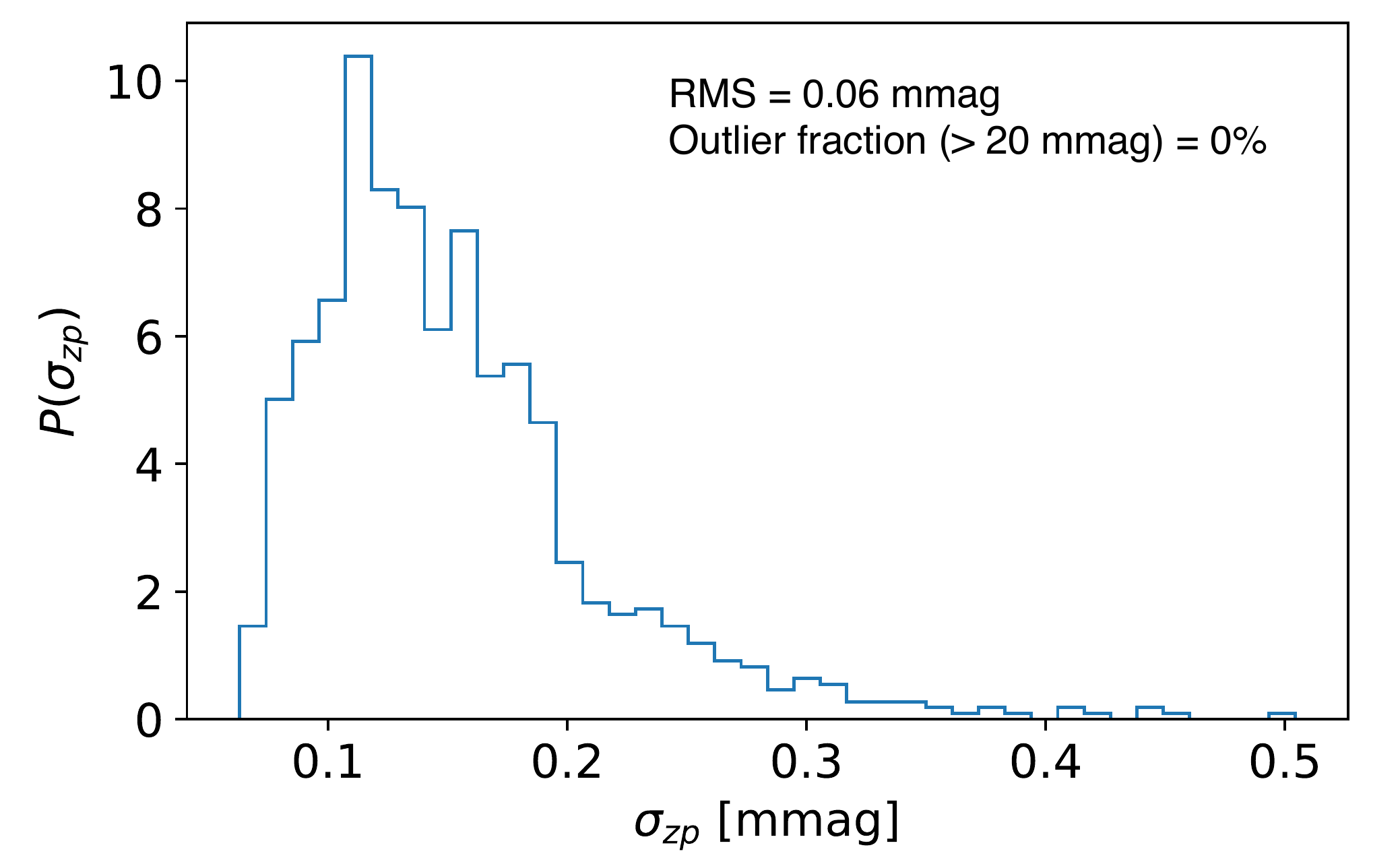}
\caption{Normalized distribution of the error in the zeropoint value, $\sigma_{zp}$ for 1000 randomly selected sensor-visits. The distribution is quite asymmetrical. The RMS is 0.06 mmag, fulfilling the requirements (15 mmag). We also find that none of the visits deviate from the median by more than 20 mmag, in compliance with the requirements.}
\label{fig:PA34}
\end{figure}

The presence of clouds, calibration problems and noise fluctuations can lead to changes in the estimated magnitude zeropoint. Accurate estimation of photometric redshifts, distance moduli to supernova, and proper separation of stellar populations require stable zeropoints across the sky. In particular these science cases require that the uncertainty in the zeropoints, $\sigma_{zp}$, should show an RMS lower than 15 mmag and no more than $20\%$ of the images should have a deviation larger than 20 mmag (see the LSST-SRD for more details). We randomly select 1000 sensor-visits and check the distribution of $\sigma_{zp}$, depicted in~\figref{PA34}. We find that the RMS of the distribution is 0.06 mmag, fulfilling these requirements by a very wide margin. This is mainly a consequence of the lack of clouds and other fluctuations that may affect the zeropoint calibration (e.g., temperature/gain fluctuations in the sensors, etc.). We also find that none of the sensors have an error in the zeropoint that deviates from the median more than 20 mmag. Finally, we compare input and output magnitudes, for both stars and galaxies using \texttt{CMODEL} magnitudes~\citep{2018PASJ...70S...5B}. We compute the median of the difference between them and obtain 17 mmag, which is smaller than the maximum allowed in the LSST-SRD (20 mmag). 

These tests show that the processing pipeline performs as needed, and that our images are generated with consistent zeropoints.

\subsection{Depth requirements}
\label{sssec:depth}
The statistical power of LSST for dark energy studies relies largely on the achieved depth of the survey. A lower depth translates to an overall smaller number density of galaxies, thus increasing the statistical error budget. A careful balance between depth and area is important since a smaller survey area increases the overall statistical uncertainty. Taking this into consideration the LSST-SRD sets a minimum acceptable per-visit image depth of 24.3 mag in $r$-band, given a fiducial sky brightness of 21 mag/arcsec$^2$, exposure time of 30 s, airmass=1 and fiducial seeing (FWHM) of 0.7 arcseconds. In order to mimic this we select the visits that fulfill the following (LSST-SRD):
\begin{itemize}
\item Altitude $> 80$ degrees.
\item $0.68\arcsec <$ seeing (FWHM) $ < 0.71\arcsec$.
\item Sky-brightness (in $r$-band) $ \geq 21$ mag. 
\end{itemize}
We obtain a total of 520 sensor-visits fulfilling these criteria. We then compute the median 5-$\sigma$ depth  using the magnitude errors (as described in \secref{masking}) and compare with the predicted depth by OpSim. After this, we check that the median of the depth distribution is deeper than the minimum required depth as defined in the LSST-SRD.

\begin{figure}
\centering
\includegraphics[width=0.85\columnwidth]{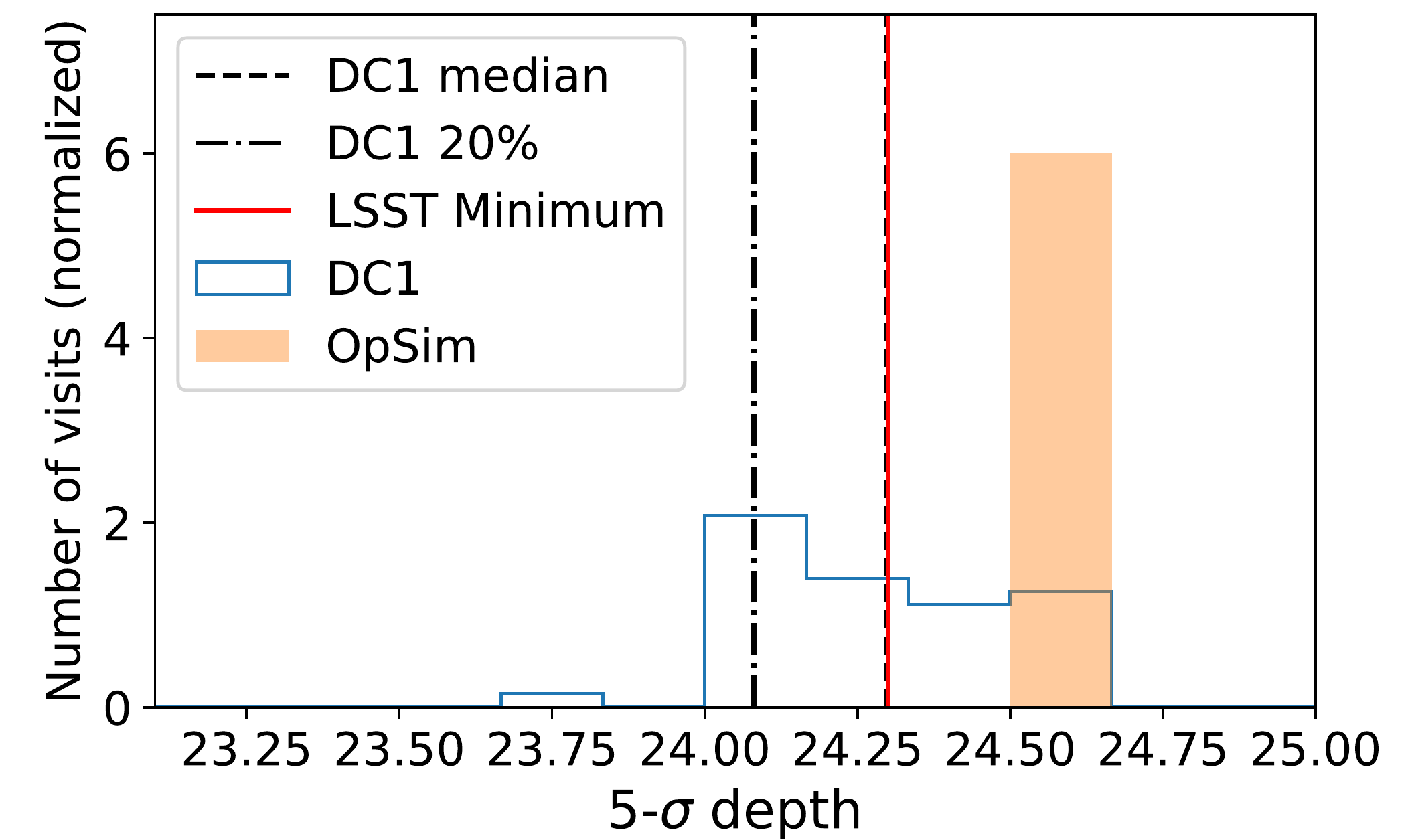}
\caption{Measured depth in visits with Altitude $>80$ deg, $0.68 \arcsec <$ seeing $ < 0.71 \arcsec$ and Sky-brightness (in $r$-band) $\geq 21$ (blue histogram) compared to the predicted depth by OpSim (solid orange histogram). The median of this distribution (dashed line) is very close to the LSST-SRD minimum depth, 24.3 (red vertical line), the 20th percentile is also shown and we can appreciate that it is larger than 24.0 as established by the LSST-SRD. All the visits fulfilling the criteria above have the same predicted depth by OpSim.}
\label{fig:DF1_checks}
\end{figure}

Clustering measurements rely on measuring changes in the number density of galaxies as a function of position to obtain cosmological information. Depth variations, usually due to changes in the sky-background levels, seeing and survey strategy (dithering), lead to variations in the observed number density and, if left unaccounted for, they can lead to biases in the inferred cosmological measurements. This is why we also check that no more than 20\% of the visits have a depth lower than 24.0 by computing the lower 20th percentile in the depth distribution. The results of these checks are depicted in \figref{DF1_checks}. We find that the median of the depth distribution in the selected visits, $24.297 \pm 0.009$ mag, is compatible with the minimum of 24.3 mag. We find as well that the 20th percentile, 24.1, is larger than the minimum. We also check that in a given visit, the variation in the field of view is within the requirements. The LSST-SRD establishes that, in a representative visit no more than 20\% of the field of view have a depth 0.4 mag. brighter than the nominal (24.3). We select visit 2218486 since its median depth is 24.3. We find that the 20th percentile is 24.29, fulfilling the criteria. This test demonstrates that we generate images with the required depth and in concordance with our inputs.

\subsection{PSF requirements}
\label{sssec:psf}
\begin{figure}
\centering
\includegraphics[width=0.85\columnwidth]{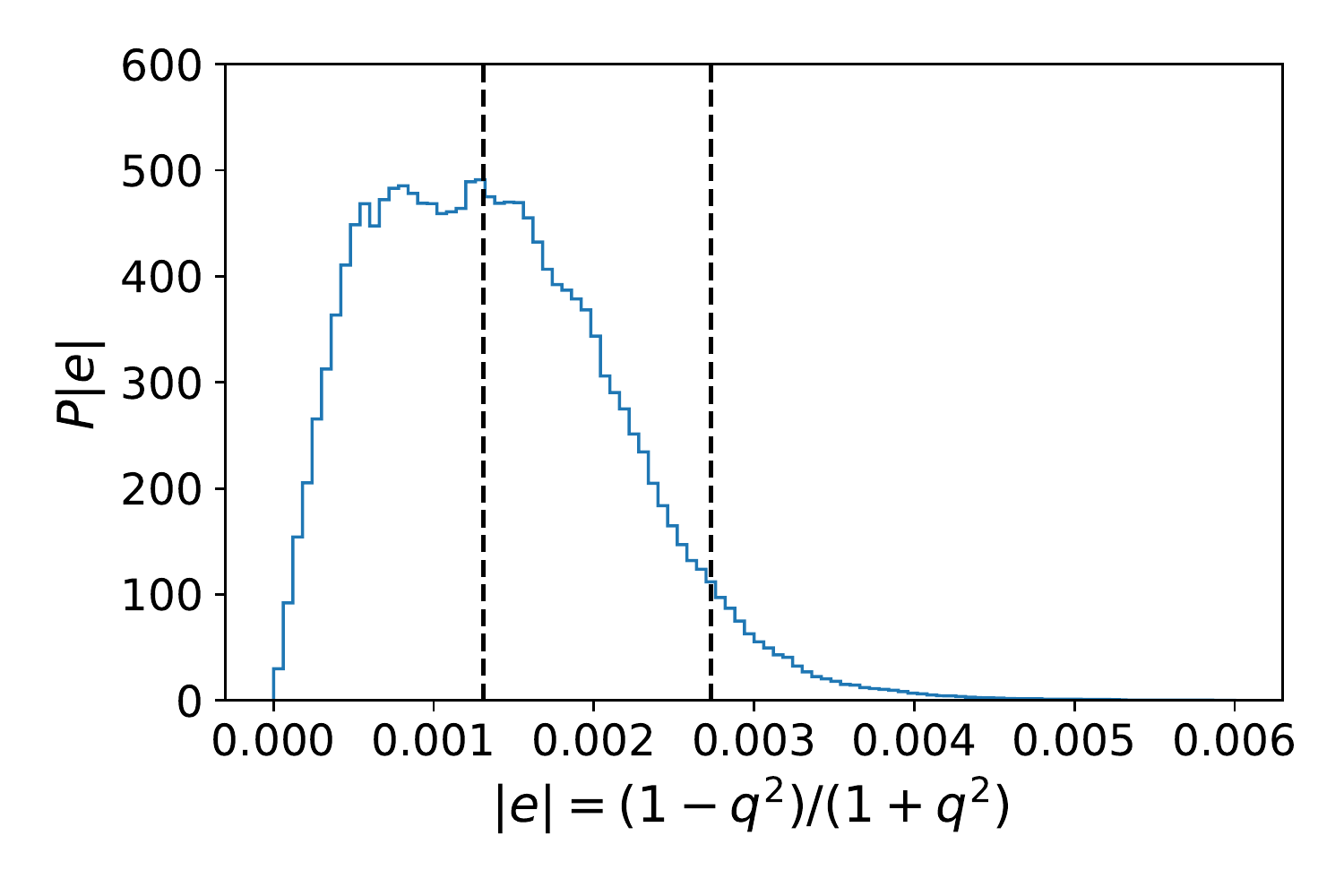}
\caption{PSF ellipticity distribution accumulated for 520 sensor-visits measured at the positions of detected objects. The median ($\approx 0.001$) and 90th percentile ($\approx 0.003$) are shown as the dashed lines. Note that these values are an order of magnitude lower than the upper limits specified in by the LSST-SRD (0.05 and 0.1, respectively).}
\label{fig:SE1_DC1}
\end{figure}

Weak lensing measurements rely on the estimation of coherent patterns in the shapes of galaxies in order to extract cosmological information. These shapes are distorted by the PSF of the system (atmosphere+telescope+detector). A PSF with a large ellipticity module, compared to a relatively circular PSF, may have the same relative residual level, however, given the greater absolute value of the ellipticity, the absolute residual level is also higher and has a larger impact in the weak lensing signal. This is why the LSST-SRD sets criteria regarding the maximum modulus of the PSF ellipticity, $|e|$, for visits with the same criteria used for the depth requirements, mostly driven by weak lensing analyses. We use the distortion definition for $|e|$~\citep{1991ApJ...380....1M}:
\begin{equation}
|e| = \frac{a^{2} - b^{2}}{a^{2}+b^{2}},
\end{equation}
where $a, b$ are the semi-major and semi-minor axes of the PSF. We test exposures with PSF-FWHM $\approx 0.69\arcsec$ and no more than 10 degrees away from zenith and check that the median ellipticity is no larger than 0.05 and that no more than 10\% of the images exceed 0.1. Our analytic (and circularly-symmetric) PSF models should, by design, fulfill these criteria. However, we must test whether the reconstructed PSF also fulfills them. The PSF was reconstructed using the PSFEx~\citep{2011ASPC..442..435B} implementation in the LSST software stack. We tested this in the processed data by using the same 520 sensor-visits used to check the depth requirements described above. We checked the modulus of the PSF ellipticity at the position of detected stars, using their measured ellipticity in these visits, and accumulating them in the histogram shown in \figref{SE1_DC1}. We obtained that the median ellipticity is $\approx 0.001$ and the 90th percentile is $\approx 0.003$, below the maximum values allowed by the LSST-SRD (0.05 and 0.1, respectively). The second data challenge (DC2) will use more realistic PSF models and we expect these margins to be different.


For weak lensing analyses, correct modeling of the PSF is crucial~\citep{2004MNRAS.353..529H} and both the LSST-SRD and the DESC-SRD specify explicit requirements about PSF residuals. In particular, the LSST-SRD states that using the full survey data the auto- and cross-correlations ($E_{1}, E_{2}, E_{X}$) of the PSF residuals over an arbitrary field of view should be below (3 $\times 10^{-5}$) for $\theta \leq 1$ arcmin, and below ($3 \times 10^{-7}$) for $\theta \geq 5$ arcmin. 

To check these criteria we calculate $E_{1}, E_{2}, E_{X}$ using the definitions of the LSST-SRD:
\begin{eqnarray}
e_{1} = \frac{\sigma^{2}_{1} - \sigma^{2}_{2}}{\sigma_{1}^{2}+\sigma_{2}^{2}},\\
e_{2} = \frac{2\sigma^{2}_{12}}{\sigma_{1}^{2}+\sigma_{2}^{2}},\\
E_{1} (\theta) = \langle \delta e^{(i)}_{1}\delta e^{(j)}_{1} \rangle,\\
E_{2} (\theta) = \langle \delta e^{(i)}_{2}\delta e^{(j)}_{2} \rangle,\\
E_{X} (\theta) = \langle \delta e^{(i)}_{1}\delta e^{(j)}_{2} \rangle.
\end{eqnarray}
The quantities $\sigma_{1}^{2}, \sigma_{2}^{2}$ are the second-order moments of a source along some set of perpendicular axes and $\sigma^{2}_{12}$ is the covariance, $\delta e_{1}, \delta e_{2}$ are the residuals, and the angle brackets indicate averaging over all pairs of stars $i$, $j$ at a given angular separation $\theta$.

\begin{figure}
\centering
\includegraphics[width=0.85\columnwidth]{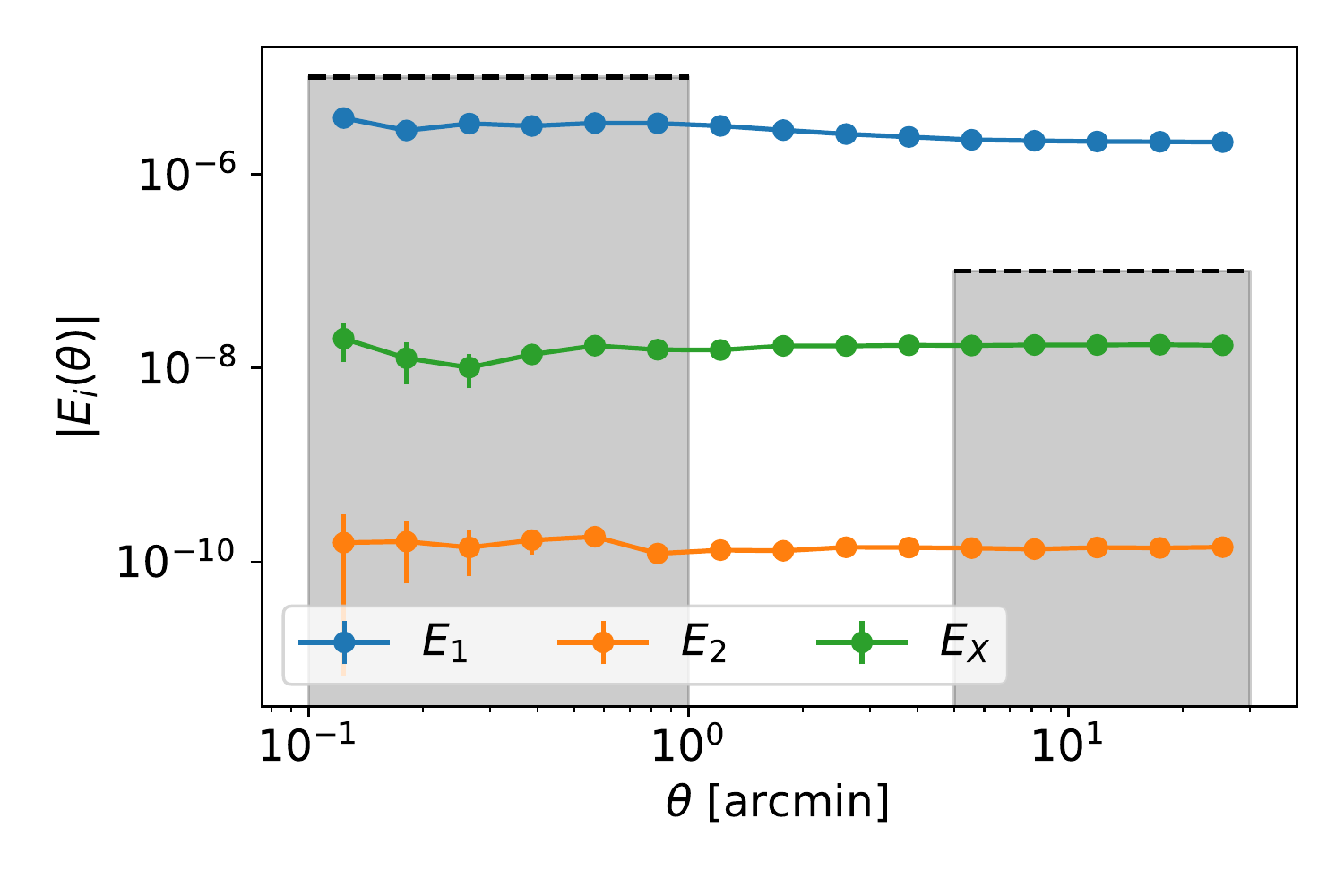}
\caption{Auto and cross-correlation functions, $E_{1}$ (blue), $E_{2}$ (orange), $E_{X}$ (green) of the PSF residuals as a function of the aperture angle $\theta$. The median value of $E_{1}$, $E_{2}$, and $E_{X}$ should be within the shaded regions in order to pass the requirements. All the correlations pass the requirements for $\theta \leq 1$ arcmin but $E_{1}$ does not meet the requirement for $\theta \geq 5$ arcmin.}
\label{fig:TEx}
\end{figure}

\begin{figure}
\centering
\includegraphics[width=0.85\columnwidth]{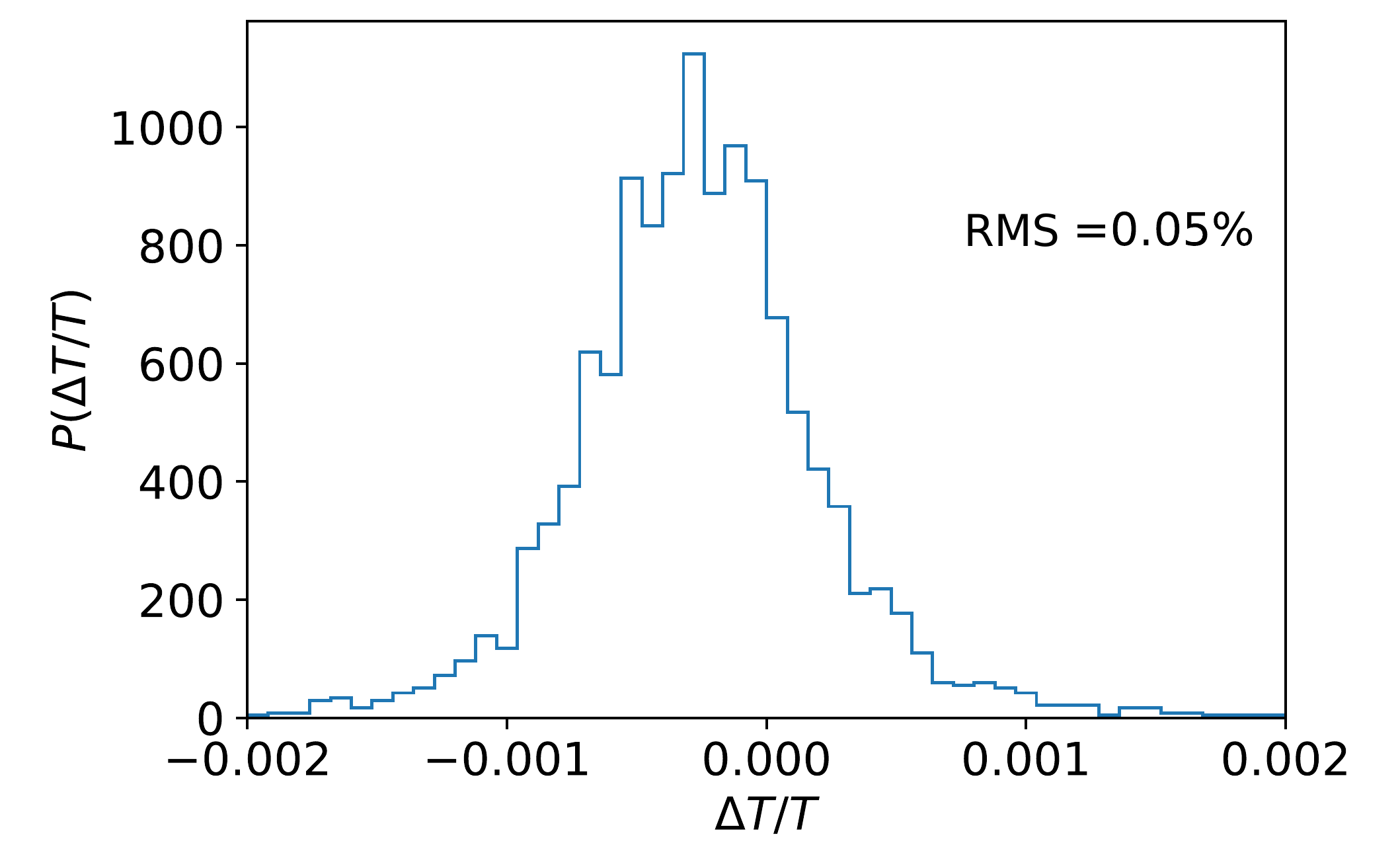}
\caption{Normalized distribution of the relative difference in the trace of the second-order moments, $\Delta T/T$ between input and output PSF. We see that the standard deviation of the distribution is 0.05\%, in compliance with the requirement (0.1\%).}
\label{fig:WL4-Y10}
\end{figure}

In practice, we compute the PSF-corrected moments of high signal-to-noise (SNR $>$ 100) stars across the field of view using \texttt{TreeCorr}~\citep{2004MNRAS.352..338J}. Our findings are shown in \figref{TEx}. We can see that $E_{2}$ and $E_{X}$ fulfill the requirements. However, $E_{1}$ does not meet the large scale requirements. This difference between $E_{1}$ and $E_{2}$ exists because $E_{1}$ is calculated in the direction of the pixels, whereas $E_{2}$ is computed in the diagonal and is subject to smaller biases. The orientation of these axes is determined somewhat arbitrarily; we could choose a different set of axes for $E_{1}$ and $E_{2}$ so both would meet the requirements. Nevertheless, PSF modeling is an area of active development within LSST Data Management and we expect improvements in future data challenges. 
We can compare the results in~\figref{TEx} to the measurements in Figure 8 of~\citet{2018PASJ...70S..25M} for $\rho_{1}$~\citep{2010MNRAS.404..350R}:
\begin{equation}
\rho_{1} = \langle (\delta e_{1}+\mathrm{i}\delta e_{2})(\delta e_{1}-\mathrm{i} \delta e_{2}) \rangle,
\end{equation}
where $\mathrm{i}$ denotes the imaginary unit. So $\rho_{1} \approx E_{1} + E_{2}$. We see that both $E_{1}$ and $E_{2}$ are smaller than the measured $\rho_{1}$ for the range of scales shown in the HSC analysis by~\citet{2018PASJ...70S..25M} ($\theta \geq 3$ arcmin). However, even in the more complex case of the HSC PSF, the residual correlations for $\theta \geq 5$ arcmin are below the requirements (the published HSC work do not show the residuals for scales below 1 arcmin).
Finally, the DESC-SRD requires that the systematic uncertainty in the PSF model defined using the trace of the second moment matrix should not exceed 0.1\% for full-depth (Y10) DESC weak lensing analysis. We randomly select 3,000 visits, obtain the input and measured PSF, and measure the trace of the second order moments, $T$ with GalSim. We then compute the relative difference, $\Delta T/T$, obtaining the results depicted in \figref{WL4-Y10}. We find that our dataset shows that the standard deviation of the distribution is 0.05\%, lower than the requirement. 

As seen in this section, our images and catalogs satisfy most of the requirements by a good margin, demonstrating our ability to both generate and process high-quality data. 

\section{Data selection and masking}
\label{sec:data_selection}
In this section, we describe how we take advantage of the fact that we have full knowledge about the simulated sources in order to get a ``clean" data sample for clustering tests. We also describe the catalog mask and how we generate maps of different observational effects (seeing, sky-brightness, etc.) present in the simulation. Unless explicitly stated, the procedures and selections made in this section are performed in both the dithered and undithered simulations.

\subsection{Sample selection}
\label{ssec:sample_selection}

As we previously mentioned, the presence of sources with poorly determined fluxes, positions or shapes can affect the clustering and lensing signals. For example, if we consider the ellipticity distribution, a small fraction of sources with unrealistically large ellipticity can significantly change the inferred variance and mean of the distributions and bias any cosmological constraints. Similarly, if we consider that the positions of bright stars are uncorrelated with the positions of galaxies, the detection of spurious sources near bright stars can lead to biased clustering statistics at all scales. These are just two of the many examples that showcase the importance of having a clean, well-understood sample to perform cosmological analyses.  

In this subsection we are going to use the \textsf{S+M} technique to identify processing flags or thresholds in variables that may allow us to get a clean sample for clustering. Given that the pipeline used for data reduction is essentially the same as the one used in~\citet{2018PASJ...70S..25M}, we could potentially perform similar cuts. However, the lack of multiband coverage and of some catalog quantities, such as the so-called blendedness, lead us to propose our own selection cuts, although we follow the criteria in~\citet{2018PASJ...70S..25M} as guidance.

The methodology to perform the selections is simple: we check the primary detected sources that have no match using the \textsf{S+M} technique and we compute the fraction of objects that are flagged, $f_{u,i} = N_{\rm{flag}_{i}, u}/N_{\rm{total}, u}$, where the subscript $u$ stands for unmatched, and compare it to the corresponding fraction of flagged matched primary detected sources, $f_{m,i} = N_{\rm{flag}_{i}, m}/N_{\rm{total}, m}$, where the subscript $m$ stands for matched, for each of the flags, flag$_{i}$, in the catalog. If the ratio $f_{u,i}/f_{m,i}$ is larger than 50 for a particular flag and $f_{m,i} < 0.01$, i.e., less than 1\% of the matched primary sources have that flag, it means that the presence of that flag is a good indicator of problematic sources. Thus we eliminate the sources with those flags. We also repeat the same procedure looking for the absence of a certain flag or whether a quantity is frequently measured as not-a-number, \texttt{NaN}.


We notice that some of the flags very efficiently distinguish unmatched from matched objects. For example, \texttt{base\_ClassificationExtendedness\_flag = True}, which means that there was a failure at the time of deciding whether a source was extended or point-like, eliminates more than 30\% of the objects with no match, while barely affecting the matched objects. Three other flags would be fairly efficient at filtering out unmatched objects but, if we were to use them, we would lose $\approx 50\%$ of our sample. These include \texttt{modelfit\_CModel\_flags\_smallShape==False} which means that the initial parameter guess did not result in a negative radius (if \texttt{True} the initial guess for the radius would be negative). Intuitively, we expect this flag to be \texttt{False} for well-behaved objects and thus, we should not use this for our selection. Another case where the fraction of flagged unmatched (and matched) objects is very large is \texttt{modelfit\_CModel\_flags\_region\_usedFootprintArea==True}, which means that the pixel region for the initial fit was defined by the area of the footprint. This flag is not necessarily indicative of problems with the measured object. Finally, we see that \texttt{modelfit\_CModel\_flags\_region\_usedPsfArea==False} also affects a very large fraction of unmatched and matched objects. These objects are such that the pixel region for the initial fit was not set to a fixed factor of the PSF area, which is not indicative of any problems with the source.

As a result, we eliminate from our sample all the sources that fulfill at least one of the following conditions:
\begin{itemize}
\item \texttt{detect\_isPrimary = False}. As discussed earlier, this means that the source has not been fully deblended or is outside of the inner region in a coadd.
\item \texttt{base\_NaiveCentroid\_flag = True}. This means that there is a general failure during the source measurement.
\item \texttt{base\_SdssShape\_flag\_psf = True}. This means that there is a failure in measuring the PSF model shape in that position.
\item \texttt{ext\_shapeHSM\_HsmSourceMoments\_flag\_not\_contained = True}. This means that the center of the source is not contained in its footprint bounding box.
\item \texttt{modelfit\_DoubleShapeletPsfApprox\_flag = True}. This means that there is a general failure while performing the double-shapelet approximation to the PSF model at the position of this source (see Appendix 2 in~\citet{2018PASJ...70S...5B} for more details).
\item \texttt{base\_PixelFlags\_flag\_interpolated = True}. This means that there are interpolated pixels in the source's footprint.
\item \texttt{base\_PixelFlags\_flag\_interpolatedCenter = True}. This means that the center of a source is interpolated.
\item \texttt{base\_PixelFlags\_flag\_saturatedCenter = True}. This means that the center of a source is saturated.
\item \texttt{base\_ClassificationExtendedness\_flag = True}. This means that there is a general failure when using the \texttt{extendedness} classifier.
\item \texttt{modelfit\_CModel\_flags\_region\_usedInitialEllipseMin = True}. This means that the pixel region for the final model fit is set to the minimum bound used in the initial fit.
\item \texttt{base\_SdssShape\_x/y = NaN}. This means that the centroid position (either in the x or y axes) is measured as \texttt{NaN}.
\item \texttt{base\_SdssCentroid\_x/yErr = NaN}. This means that the error in the centroid position (either in the x or y axis) is measured as \texttt{NaN}.
\end{itemize}

After these cuts we keep 8.25 million objects in the dithered catalog and 7.51 million objects in the undithered catalog. We will refer to this sample as the \textit{benchmark sample}. After the cuts, the ratio of unmatched objects decreases by $\approx 60\%$ from $\sim 5\%$ (of the catalog before cuts) to $\sim 3\%$ (of the benchmark sample), while we retain $99.7\%$ of the matched objects. In DC1 we are limited by only having $r$-band information. The addition of other imaging bands will provide independent information that will decrease the fraction of noise fluctuations that make it into the catalog and will allow selection of an even cleaner sample (e.g., by performing selection cuts in color-color diagrams).

\begin{figure}
\centering
\includegraphics[width=0.9\columnwidth]{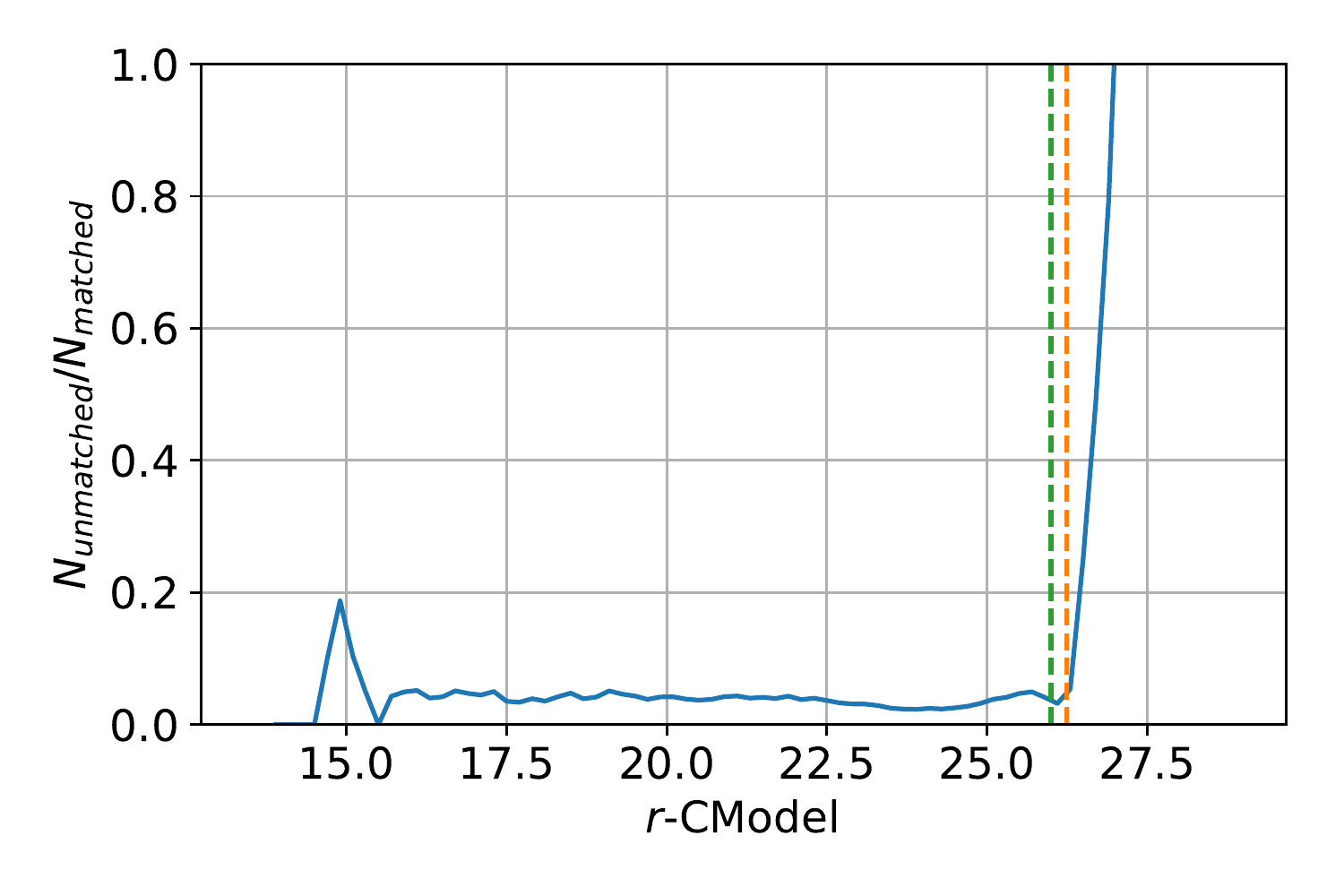}
\includegraphics[width=0.9\columnwidth]{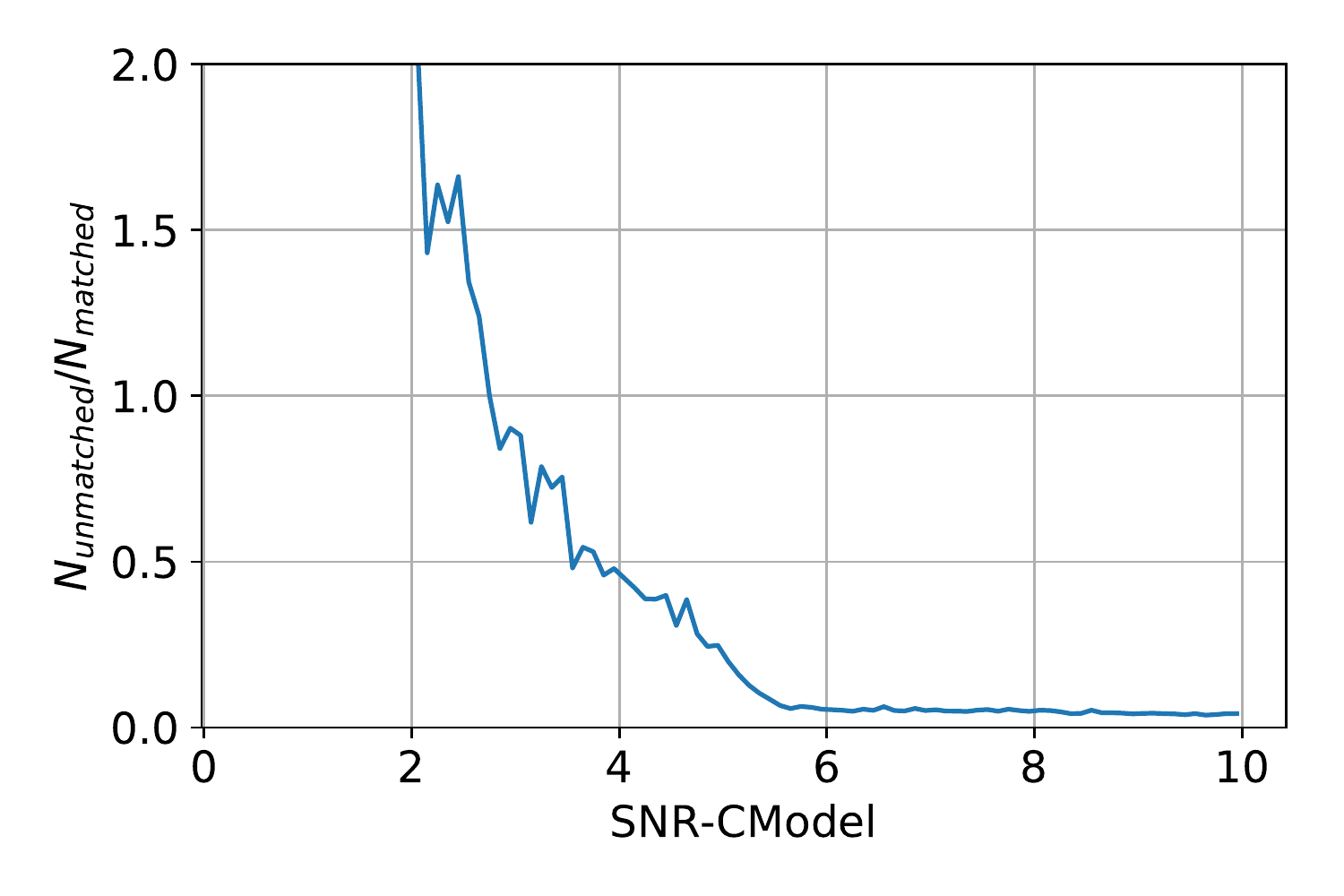}
\caption{Top: Ratio of unmatched to matched sources in the benchmark sample as a function of magnitude. The dashed vertical lines show the median depths for the dithered (orange) and undithered (green) fields. Bottom: Ratio of unmatched to matched sources in the benchmark sample as a function of SNR.}
\label{fig:snr_mag_selection}
\end{figure}

We now focus on how many of these objects are matched as a function of magnitude and signal-to-noise ratio. In \figref{snr_mag_selection}, we can see that the fraction of unmatched objects grows very quickly for $r > 26$, and for SNR$<6$. Therefore, $r < 26$ and/or SNR$>6$ appear to be sensible selection criteria to ensure good quality data. The peak at $r \sim 15$ is likely a consequence of bright objects that have been clipped in the simulation process to emulate saturation; these objects have measured magnitudes of $r \sim 15$, however, their true magnitudes are brighter by one or more magnitudes and thus, appear as unmatched by the \textsf{S+M} matching strategy. 

We select our \textit{clustering sample} using the following criteria:

\begin{itemize}
\item Full set of cuts of the benchmark sample.
\item $20 \leq$ \texttt{r\_mag\_CModel} $\leq 25.5$. 
\end{itemize}
Note that we do not use the SNR$>6$ criterion, this is because the magnitude cut nearly guarantees it.
These selection cuts ensure a low fraction of unmatched objects and also, as we will justify in following sections, high purity of our galaxy sample (low stellar contamination) once we select galaxies via the \texttt{base\_ClassificationExtendedness\_value} classifier.

\subsection{Star/galaxy classification}
\label{sec:sg_sep}
\begin{figure}
\centering
\includegraphics[width=0.9\columnwidth]{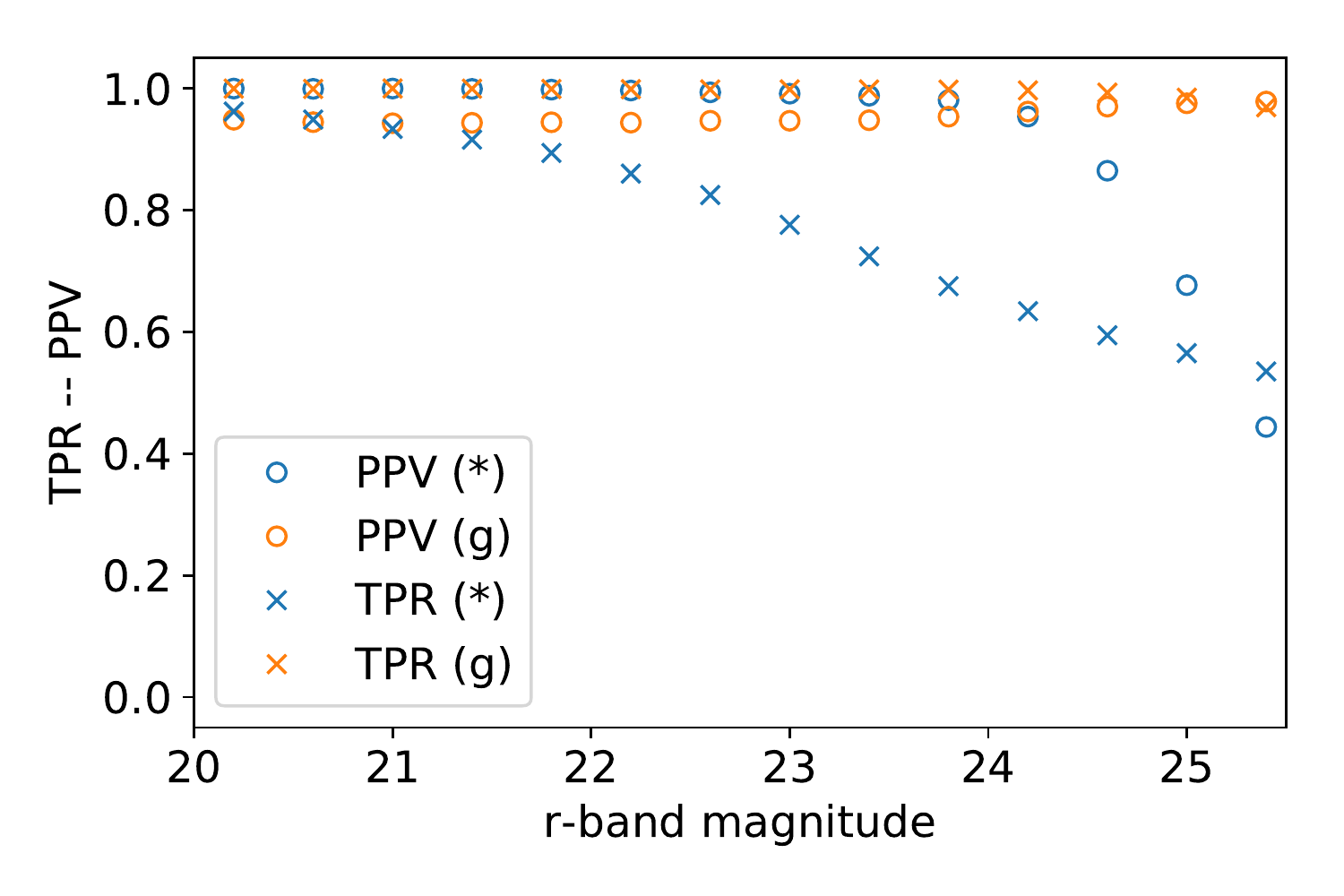}
\caption{Performance of the \texttt{base\_ClassificationExtendedness\_value} as star-galaxy classifier as a function of magnitude. This classifier achieves a high galaxy purity and completeness. We show the true positive rate (TPR) for stars (blue crosses) and galaxies (orange crosses) and the positive predictive value (PPV) for stars (blue open circles) and galaxies (orange open circles). Both the TPR and PPV for galaxies are high in the range of magnitudes that we are studying.}
\label{fig:star_galaxy_separation}
\end{figure}

For weak lensing and clustering analyses, it is important to have a pure galaxy sample and good control over the fraction of stars that are classified as galaxies. Our pipeline includes the variable \texttt{base\_ClassificationExtendedness\_value} (see~\citet{2018PASJ...70S...5B} for more details), which we will refer to as \texttt{extendedness}, and can be used as a proxy to separate stars from galaxies as shown in~\citet{2018PASJ...70S..25M} and~\citet{2018PASJ...70S...5B}. In this work we say that an object has been classified as a galaxy if \texttt{extendedness=1}, and that the object has been classified as a star if \texttt{extendedness=0}. To evaluate the performance of \texttt{extendedness} as star/galaxy classifier in DC1, we use the clustering sample in the dithered field, although we find similar results using the undithered field, and match it to our input catalog using the \textsf{S+M} method. After that, we follow~\citet{2018MNRAS.481.5451S} and compute the true positive rate (TPR), usually referred to as \textit{completeness}, and positive predictive value (PPV), usually called \textit{purity}, defined as:
\begin{eqnarray}
\mathrm{TPR} = \frac{\mathrm{True Positive}}{\mathrm{True Positive}+\mathrm{False Negative}}, \\
\mathrm{PPV} = \frac{\mathrm{True Positive}}{\mathrm{True Positive}+\mathrm{False Positive}}.
\end{eqnarray}
For galaxies, true positives are objects classified as galaxies and matched to galaxies; false negatives are objects classified as stars but matched to galaxies; and false positives are objects classified as galaxies but matched to stars. This way, we know the total stellar (or galaxy) contamination as a function of measured magnitude. The results are depicted in \figref{star_galaxy_separation}. We see that at the fainter end ($r \approx 25$), the PPV (purity) of the stellar sample using the \texttt{extendedness} classifier starts to decrease, getting as low as 50\% for the last bin in our analysis, while the PPV for the galaxy sample remains stable across the selected range of magnitudes. We obtain a total $f_{star}=1.4\%$. For a more restricted magnitude threshold of $r < 25$ we can get $f_{star} = 0.7\%$. Note that these fractions are larger than those presented in~\citet{2018PASJ...70S...5B}. This is primarily because of our broader PSF, making the \texttt{extendedness} classifier perform a little bit worse, and that we do not include any cuts on resolution. However, this level of stellar contamination is acceptable for the purposes of this work. Note that this selection focuses on galaxy purity and completeness and should be modified for stellar studies.

\subsection{Depth maps and footprint masking}
\label{sec:masking}

In order to estimate the depth in the coadd catalogs we generate a map of our footprint in flat-sky approximation (i.e., using the plate carr\'{e}e projection), with resolution of $1.74$ arcminutes, containing the sources in the \textit{benchmark sample}. This resolution allows us to accurately estimate the power spectra up to $\ell \sim 6000$, where the power spectra will be mostly dominated by shot-noise. Then, for each cell in the map, we bin the objects in magnitude and compute  the median SNR in each bin, after this we find the magnitude bin closest to SNR=5, using \texttt{CMODEL} fluxes and their quoted uncertainties.

\begin{figure}
\centering
\includegraphics[width=0.85\columnwidth]{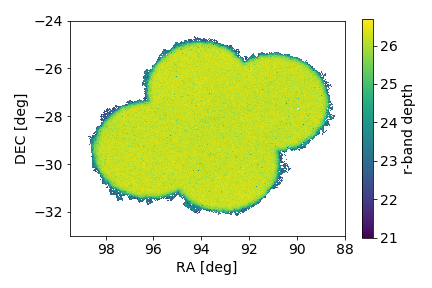}
\includegraphics[width=0.85\columnwidth]{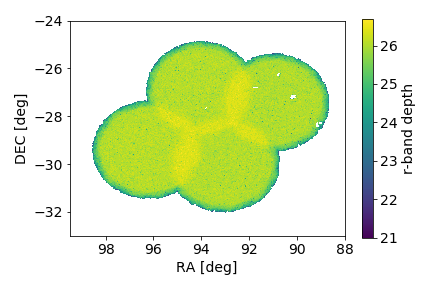}
\includegraphics[width=0.85\columnwidth]{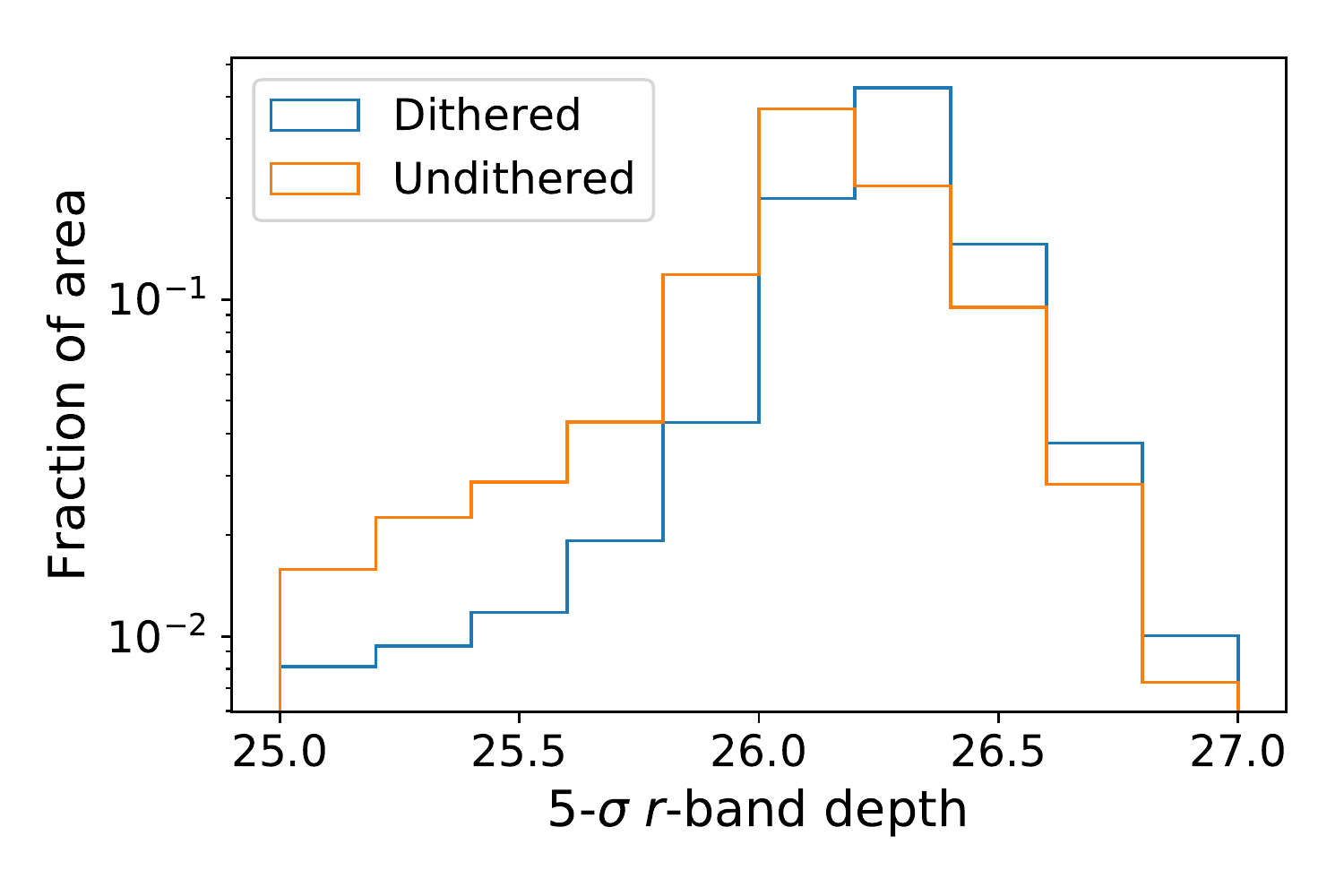}
\caption{5-$\sigma$ depth maps for the dithered (top) and undithered (middle) fields. There is an increased depth in the overlapping parts of the LSST field of view in the undithered field but the median depth is lower. We see some holes in the undithered footprint due to missing data. The maps have a resolution of $1.74$ arcmin. We also show the 1D distributions of depth (bottom) for both fields for easier comparison.}
\label{fig:depth_maps}
\end{figure}

These maps are shown in \figref{depth_maps}. We can see that the dithered simulation is indeed very uniform ($> 50\%$ of its footprint lies in the same depth bin) showing the success of the dither strategy. Additionally, we can see that the undithered simulation has a smaller median depth.

\begin{figure}
\centering
\includegraphics[width=0.9\columnwidth]{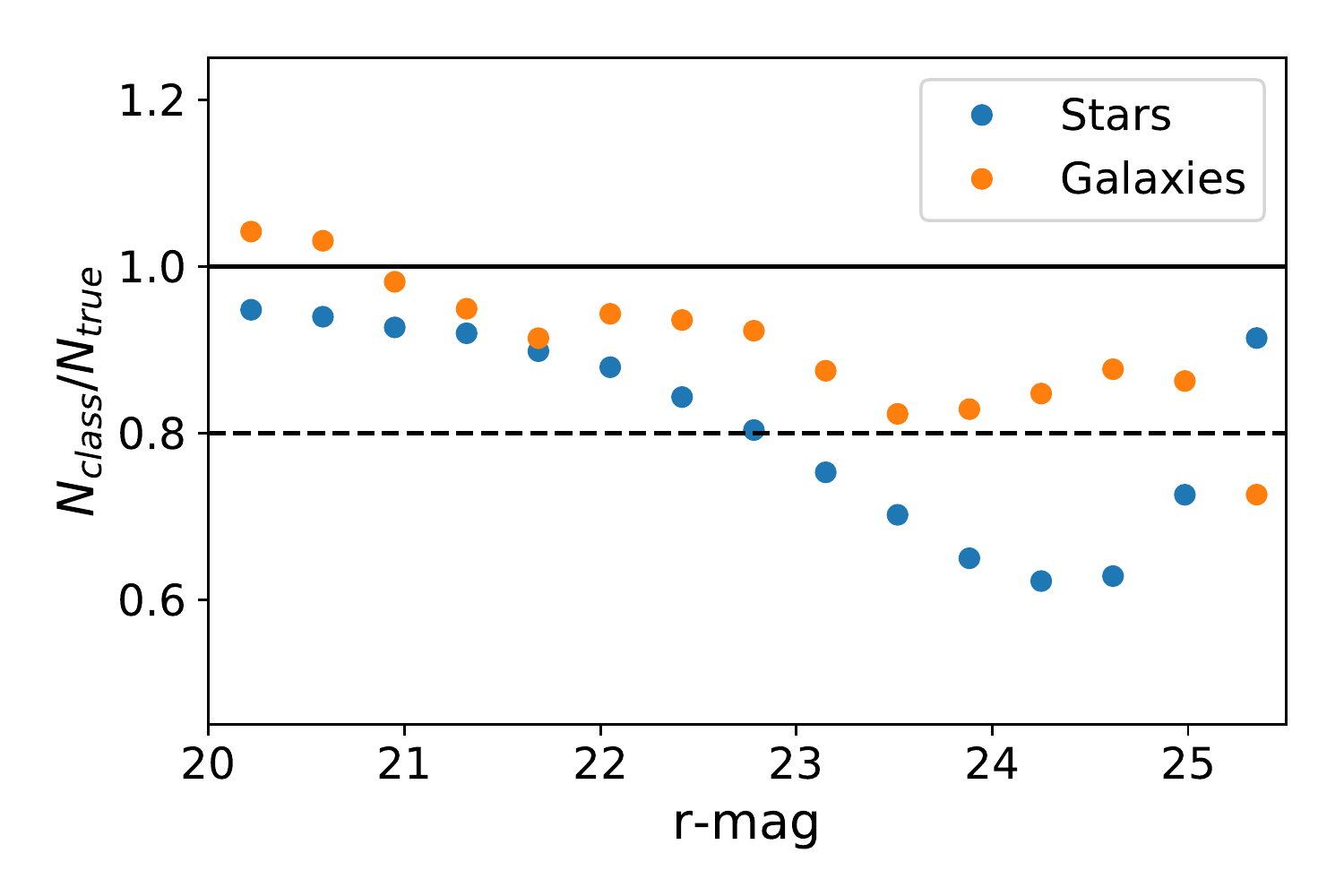}
\caption{Ratio of number of objects in the clustering sample classified as stars and number of input stars as a function of magnitude. We also compute this same ratio for galaxies. This is basically a measurement of the detection efficiency (completeness) of stars and galaxies in the clustering sample. Note that the ratio can go above one since there will be stars classified as galaxies (and galaxies classified as stars) and artifacts that will pass our cleaning cuts.}
\label{fig:stellar_detection_efficiency}
\end{figure}

We also check the depth by computing the detection efficiency (completeness\footnote{Note that completeness is defined differently in the star/galaxy classification section~\ref{sec:sg_sep}.}) of stars and galaxies as a function of magnitude. To do so, we use the objects in the input catalog and select those that lie within the simulated footprint. After this, we compute the number of detected objects in the \textit{clustering sample} classified as stars and classified as galaxies as a function of their \texttt{CMODEL} magnitude, and divide by the number of stars and galaxies in the input catalog as a function of the true magnitude. The results can be seen in \figref{stellar_detection_efficiency}. We see that there is a high detection efficiency for galaxies $> 80\%$ up to $r \approx 25.5$.  

Given the results in the two previous subsections and this subsection, we decide to use only the galaxies that lie in cells with limiting magnitude $r \geq 25.5$ and that have been visited at least 92 times, which corresponds to 50\% of the nominal full-depth number of visits for the full 10-year LSST~\citep{Overview}. On top of that, we select those objects with magnitudes in the range $20 \leq r \leq 25.5$. This cut ensures high detection efficiency ($>80\%$) and it allows us to eliminate most of the spurious detections in the sample (2.8\%). In addition it results in a low stellar contamination ($f_{star} \approx 1.4\%$). After these cuts and with selection of objects with \texttt{extendedness=1}, we obtain 4.5 and 4.0 million objects for the dithered and undithered fields respectively. This selection cut, however does not change the fraction of unmatched objects explored in the previous subsection.

\subsection{Bright star masking}

Bright objects produce significant effects in an image that affect the detection and measurement of neighboring objects. Some examples of these effects include saturation, large diffraction spikes (not included in our simulations), scattered light (also not included in our simulations), obscuration of neighboring sources. Masking regions around these sources creates a more complicated footprint but greatly simplifies the analysis of systematic effects. In order to avoid possible biases by masking bright galaxies we will only analyze the impact of bright stars on the nearby detected objects.

\begin{figure}
\centering
\includegraphics[width=0.9\columnwidth]{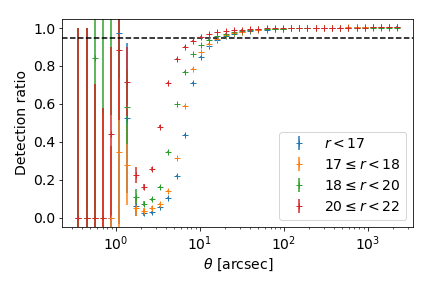}
\caption{Ratio of the median number of primary detected objects neighboring a star in a certain magnitude range in the input catalog to the median number of objects detected near any star in the input catalog, as a function of the distance to the star $\theta$. Different colors represent different magnitude ranges for the stars in the input catalog considered.}
\label{fig:bright_object_masking}
\end{figure}

\begin{figure}
\centering
\includegraphics[width=0.9\columnwidth]{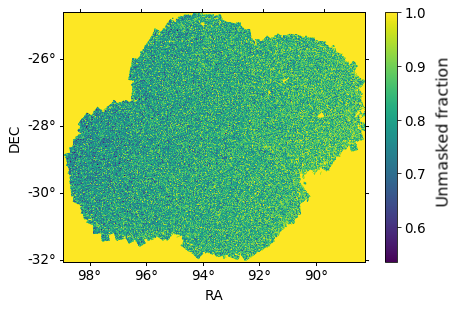}
\caption{Map showing the unmasked fraction in each pixel. We use a high-resolution ($\approx 6.4 \arcsec$) map to mask around bright stars ($r < 22$) and then we down-sample the map to a lower resolution ($\approx 1.7 \arcmin$) and remove the pixels where the masked area near bright stars is higher than 75\% of the pixel. We weight the galaxy counts in each pixel of the map by the inverse of this mask fraction to compensate for the area loss.}
\label{fig:bo_mask}
\end{figure}

In order to evaluate the effect of bright star masking, we follow the procedure described in \citet{2018PASJ...70S...7C}. Using the positions of bright objects classified as stars (\texttt{base\_ClassificationExtendedness\_value==0}) that lie within the considered footprint, and with input magnitudes in the range $m_{1} < r < m_{2}$, we count all objects from the clustering sample within a given radius $\theta$ and compute the average number of neighbors, $N_{neighbors}$. We repeat this for different radii and magnitude ranges ($ r < 17; 17 \leq r < 18; 18 \leq r < 20; 20 \leq r < 22$). Finally, we repeat this process for all stars in the input catalog in the footprint and compute $N_{neighbors, tot}$ and compute the ratio $N_{neighbors}/N_{neighbors, tot}$. This ratio is depicted in \figref{bright_object_masking} as a function of distance. We then generate the bright star mask as follows:
\begin{enumerate}
\item In \figref{bright_object_masking} we identify the radius $\theta = r_{mask, fit}$, at which $N_{neighbors}/N_{neighbors,tot}=0.95$ for each magnitude range.
\item We mask around bright stars using the value of $r_{mask, fit}$ corresponding to their magnitude, creating a high-resolution mask (6.4\arcsec-side pixels).
\item We downgrade the resolution of this mask to $\approx 2$ arcmin and eliminate pixels that are more than 75\% masked. This results in an area loss of $\approx 13\%$. 
\end{enumerate}
The resulting mask can be seen in \figref{bo_mask}.


\subsection{Blending}
As previously mentioned, our output catalogs do not include any estimates of overlap between sources, or \textit{blendedness}~\citep{2018PASJ...70S...5B}. Highly-blended objects are more likely to have biased estimations of the centroid positions, shapes, and fluxes. This can lead to overall biases in the estimated photometric redshifts and cosmological parameters. Given that the mean seeing in DC1 is larger than in HSC, the impact of blended objects will be larger. \citet{2018PASJ...70S..25M} mitigated the impact of blended objects by removing objects with more than $43\%$ of their flux in their footprint coming from overlapping sources (blendedness $ >10^{-0.375}$), which affects only $1\%$ of the objects; we expect this number to be larger for the DC1 simulations. Using dedicated image simulations from S\'{a}nchez et al., in prep. with a seeing similar to the seeing in DC1 (1.04\arcsec), in $r$-band, we find that if we select objects with $r < 25.5$ and SNR $\geq 1$, the fraction of objects with blendedness $ >10^{-0.375}$ is $\approx 6.3\%$. If we raise the minimum SNR threshold to 6, this fraction is lowered to $\approx 2.6\%$. This means that our sample will have a fraction of these objects anywhere in the range (2.6\% -- 6.3\%) but closer to 2.6\% since the fraction of objects with SNR $\leq 6$ is $\approx 0.3\%$. In any case, we do not expect that the inclusion of these objects in our two-point measurements affects the range of scales that we consider in this work. However, blending affects photometric redshifts, potentially biasing them. For DC1 photometric redshifts are not available (since we only have one band). A more rigorous study of the impact of blending on small-scale clustering measurements, and photometric redshifts is beyond the scope of the current work.

\section{Two-point clustering results}
\label{sec:results}

In previous sections we have focused on how to select a galaxy sample clean enough to perform a clustering analysis, our clustering sample. Now we want to use this sample to answer two questions: (i) is this sample actually good enough to perform a clustering measurement? (ii) given that reducing the number of objects in our catalog in order to have a cleaner sample results in an increased statistical uncertainty, we would like to understand if these cuts are actually necessary or can be relaxed. We answer these questions by measuring the angular power spectra of two different samples: (i) our so-called ``clustering sample", (ii) a ``polluted" sample, where we relax some of the selection criteria. Beyond these questions, these clustering measurements help us verify the simulation and the clustering analysis pipelines.  

\begin{figure*}
\centering
\includegraphics[width=0.30\textwidth]{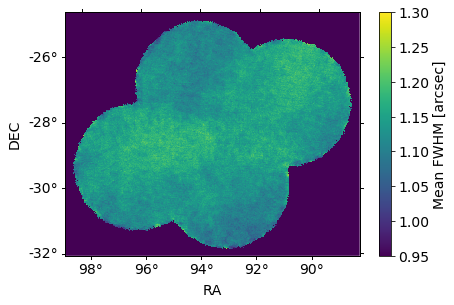}
\includegraphics[width=0.30\textwidth]{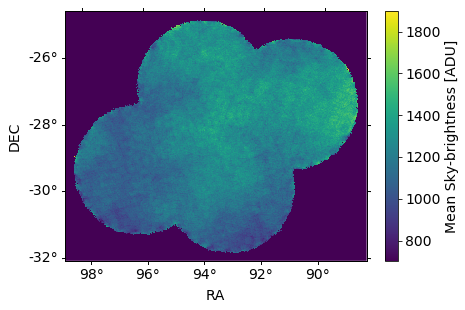}
\includegraphics[width=0.30\textwidth]{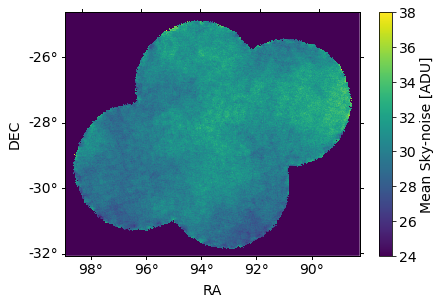}
\includegraphics[width=0.30\textwidth]{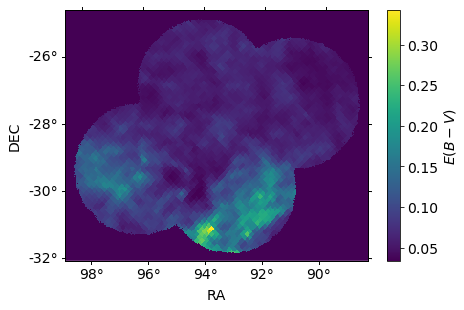}
\includegraphics[width=0.30\textwidth]{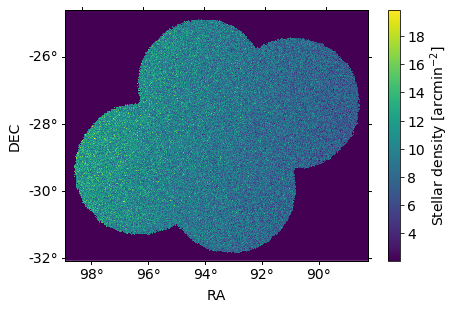}
\includegraphics[width=0.30\textwidth]{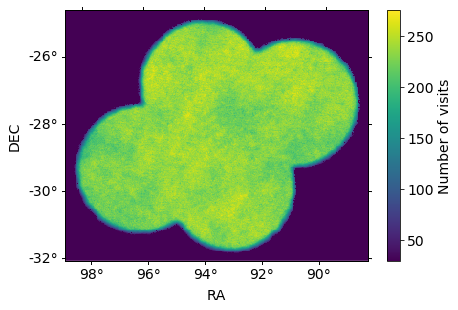}
\caption{Maps showing the different foregrounds considered in our analysis of the dithered field. From top left to bottom right: Mean PSF FWHM, mean sky-brightness, mean sky-noise, mean extinction, stellar density and number of visits in each pixel in the flat-sky maps with the same resolution as the depth maps in \figref{depth_maps}. We only show their values in the regions where the 5-$\sigma$ $r$-band depth is larger than 25.5.}
\label{fig:systematic_maps}
\end{figure*}

\begin{figure*}
\centering
\includegraphics[width=0.30\textwidth]{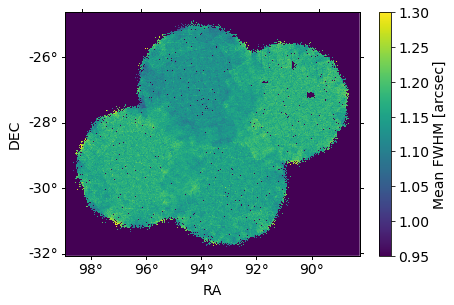}
\includegraphics[width=0.30\textwidth]{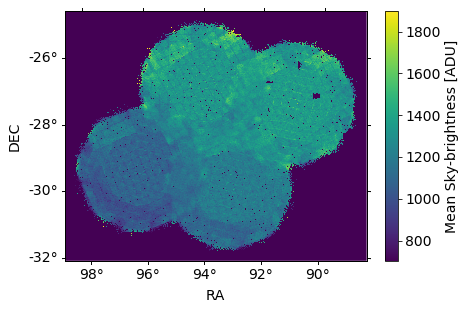}\\
\includegraphics[width=0.30\textwidth]{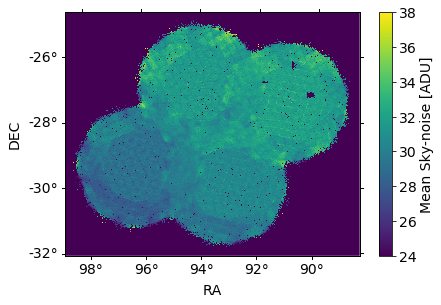}
\includegraphics[width=0.30\textwidth]{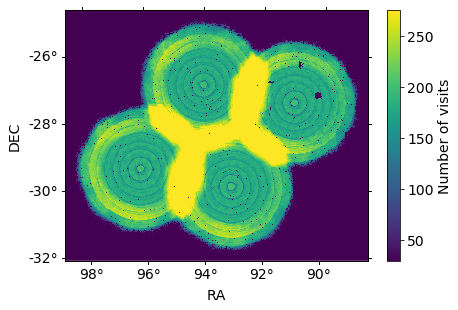}
\caption{Same as \figref{systematic_maps} for the undithered dataset. From top left to bottom right: Mean PSF FWHM, mean sky-brightness, mean sky-noise and number of visits. The maps are only shown in regions where the 5-$\sigma$ $r$-band depth is larger than 25.5. Note that we use the same extinction and stellar density maps changing the geometry of the mask since these are not affected by dithering. The circular pattern is due to the fact that the field rotates slightly in the undithered visits.}
\label{fig:systematic_maps_ud}
\end{figure*}

Apart from analyzing the impact of selection in our two-point clustering results, we also study how different observing conditions correlate with the observed number density in our sample, and whether the dithering strategy used in the simulation mitigates these correlations. We are going to consider maps of the following observing conditions:
\begin{itemize}
\item Extinction: The CatSim catalog provides the value for the magnitudes corrected for extinction using the map from \citet{1998ApJ...500..525S}, which we refer to as the SFD map.
\item Stellar contamination: In this case, we build a flat-sky map with all stars in the input catalog.
\item Sky-background/Sky-brightness: We use the observed background level in each exposure and assign that value to the pixels in the flat-sky map that lie within that exposure. After this we calculate the mean value in each pixel to build the map with the same resolution as the mask ($\approx 2$ arcmin). 
\item Sky-noise: We use the observed noise background level in each exposure and proceed as in the previous case to build a map.
\item Seeing: We proceed as before and use the observed seeing in each exposure and build a map.
\item Number of visits: We count the number of exposures overlapping with each pixel of our flat-sky maps.
\end{itemize}
These maps are shown in Figures~\ref{fig:systematic_maps} and \ref{fig:systematic_maps_ud}. We see that the spatial distributions of the different observing conditions are very different between the two simulations, even though the ranges in each of the observing conditions are very similar. 

\begin{figure}
\centering
\includegraphics[width=0.9\columnwidth]{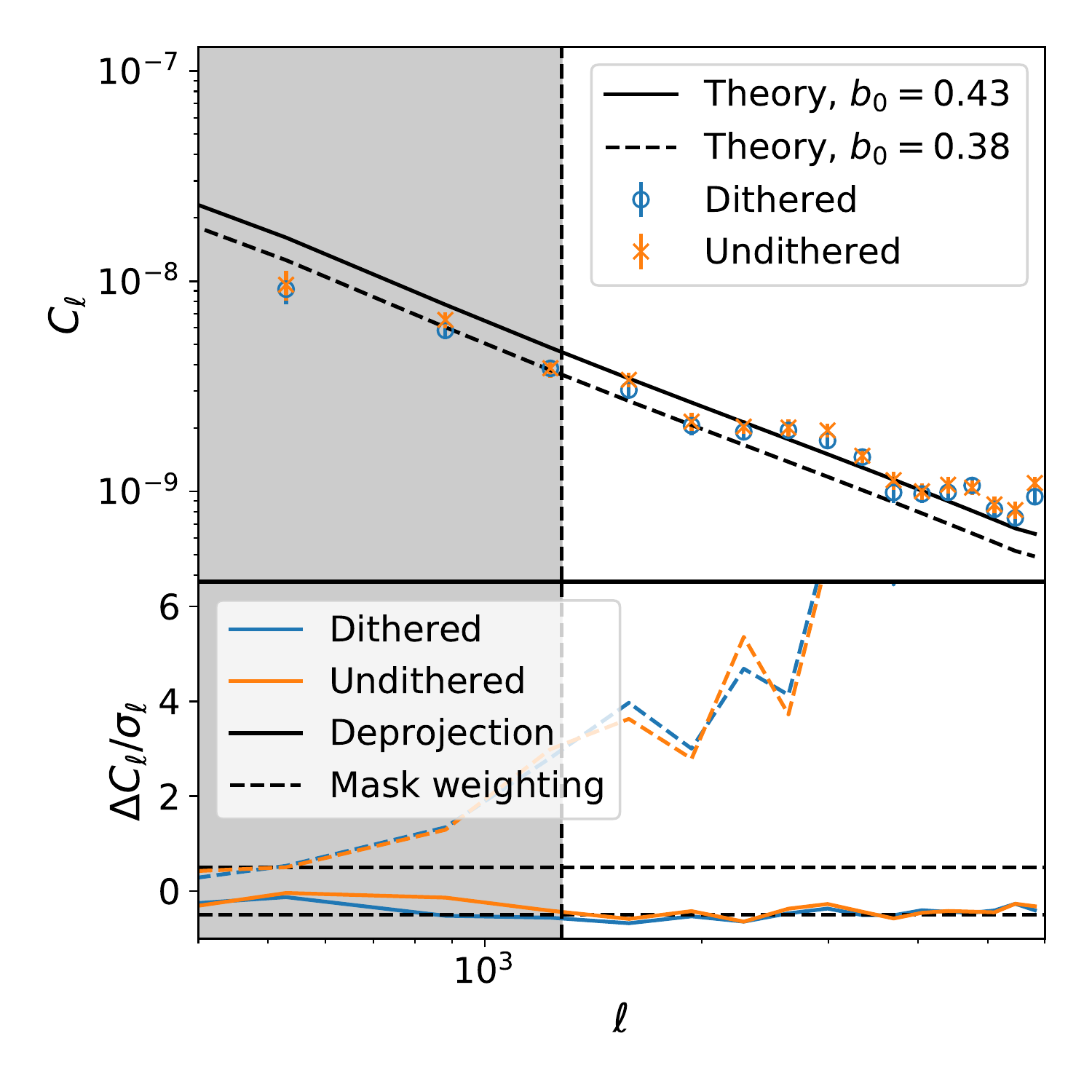}
\caption{{\bf Top panel:} Measured power spectra for undithered (orange $\times$) and dithered (open blue circles) datasets. These power spectra include the correction due to systematics using mode deprojection with \texttt{NaMaster}. The error bars are computed using jackknife. A theoretical prediction calculated with \texttt{CCL} is shown as the solid black line to demonstrate the overall qualitative agreement between the measurements and the data, we show the best-fit bias results using $\ell < 1/f_{sky}$ only ($b_{0}=0.38$, dashed line), and using $\ell > 1/f_{sky}$ only ($b=0.43$, solid line). The vertical black dashed line corresponds to $\ell = 1/f_{sky}=1192$ ($k \sim 0.9$ Mpc$^{-1}$ at the mean redshift of the clustering sample $\bar{z} \approx 1.51$) which is the minimum $\ell$ that we are sensitive to given our footprint size. {\bf Bottom panel:} Size of the correction in the power spectra, $\Delta C_{\ell}$, relative to their uncertainty, $\sigma_{\ell}$, due to deprojection (solid lines) of different observing conditions, and the correction due to weighting by the bright star mask (dashed lines) for the dithered (blue) and undithered (orange) simulations. We see that the largest impact comes from the presence of bright objects, and that it is important to account for the area lost by masking via weighting. The horizontal dashed lines correspond to a correction of 50\% of the statistical uncertainty, to provide visual guidance. The shadowed area in both panels corresponds to the region where $\ell < 1/f_{sky}$ which is not used for our analysis.} 
\label{fig:power_spectra}
\end{figure}

The power spectra computation is performed using \texttt{NaMaster}~\citep{2019MNRAS.484.4127A}. The systematics correction is also performed with \texttt{NaMaster} via mode deprojection~\citep{2016MNRAS.456.2095E,2019MNRAS.484.4127A}, which assumes that there is a linear dependence between the observed number density of galaxies and the contaminants. For our study, we compute the density contrast, $\delta_{g, i}$, in a given cell/pixel, $i$, of our maps as follows:
\begin{enumerate}
\item Ignoring the area lost due to the presence of nearby bright stars, i.e., $\delta_{g, i} = \rho_{i}/\langle \rho \rangle - 1$.
\item Compensating for the area lost due to the presence of nearby bright stars, i.e., $\delta^{*}_{g, i} = \rho/w_{i}/\langle \rho/w_{i} \rangle - 1$.
\end{enumerate}
where $\rho_{i}$ is the density in a given pixel of the map, $w_{i}$ is the inverse of the completeness mask, and $\langle \cdots \rangle$ represent ensemble averages. We choose $\Delta \ell = 352$ and compute the power spectra in the range $0 \leq \ell \leq 6000$. This choice for $\Delta\ell$ is not optimal for cosmological analyses, but it gives us a reasonably large number of bandpowers to visually check the estimated power spectra. The results for the power spectra are shown in~\figref{power_spectra}, where we can see that both the dithered and undithered catalogs yield similar results. The error bars shown and corresponding covariances are estimated using two complementary methods: On one hand by computing the Gaussian covariance with \texttt{NaMaster}, and on the other hand by considering 155 jackknife equal-area regions in our footprint. We find that both approaches give similar results and we choose to use the results from the jackknife computation.

In~\figref{power_spectra} we also compare with the theoretical prediction for the power spectra computed with \texttt{CCL}~\citep{2019ApJS..242....2C} as follows:
\begin{equation}
\begin{split}
C_{\ell}^{\rm TH} = \frac{2}{2\ell+1}\int{dz} \left(\frac{dn(z)}{dz}\right)^{2} b^{2}(z) H^{2}(z) &\times& \\
                               P\left(k = \frac{\ell+1}{r}, z\right),
\end{split}
\end{equation}
where $P(k,z)$ is the theoretical power spectrum, $b(z)$ is the galaxy bias and $\frac{dn}{dz}$ is the number density as a function of redshift, and $H(z)$ is the Hubble parameter. In particular, we use the Halofit~\citep{2012ApJ...761..152T} power-spectrum with the Millenium cosmological parameters~\citep{2005Nature.435.629S} ($\Omega_{m}=0.25$, $\Omega_{b}=0.045$, $\Omega_{\Lambda}=0.75$, $n=1$, $\sigma_{8}=0.9$, $h=0.73$), and the $dn/dz$ obtained using the true redshifts of galaxies matched in the input catalog. We use a bias, $b(z) = b_{0}/D_{+}(z)$, inversely proportional to the linear growth factor~\citep{1980lssu.book.....P}, $D_{+}(z)$, and see that, qualitatively speaking, there is a good agreement between the measurements and the prediction, as shown in~\figref{power_spectra}. We obtain a best-fit value for $b_{0}=0.43$ using the scales $\ell > 1/f_{sky}$, which are well into the non-linear regime ($k \gtrsim 0.9$ Mpc$^{-1}$ at the mean redshift of the clustering sample $\bar{z} \approx 1.51$). Additionally, we obtain the best-fit bias value, $b_{0}=0.38$, for the scales $\ell < 1/f_{sky}$, where this linear bias model is expected to be more accurate. The low value obtained for the galaxy bias (in both regimes) is a consequence of the addition of randomly positioned (i.e., non-clustered) galaxies at the fainter end ($r > 25$) in order to match the expected number density for LSST (Connolly, private comm.) that we mentioned in~\secref{inputs}. 
In any case, we do not expect to be able to fully describe the measured power spectra, given the highly nonlinear nature of the scales considered in our analysis.  Unfortunately, due to the limited of area we cannot further explore the differences between the linear and non-linear regimes, but we expect to be able to study the non-linear regime in future data challenges.

In addition, we also see in~\figref{power_spectra} that the overall impact of the systematics, evaluated as the difference between the estimated power spectra using mode deprojection and without mode deprojection, is smaller than the statistical uncertainty, $\sigma_{\ell}$, (about 50\% the size of $\sigma_{\ell}$) and that they similarly affect both dithered and undithered simulations. We do not find any statistically significant difference between the correction due to systematics for the dithered and undithered simulations. This is a consequence of several factors including: (i) the conservative cuts that we impose on our data to ensure well-behaved clustering statistics; (ii) that we only deproject using the mean value for the different observing conditions; and the lack of effects, such as vignetting, present in real images. For example, vignetting would affect the number of detected objects close to the edges of the focal plane in the undithered simulation, reducing the uniformity of the survey. However, this effect would be uniform across the footprint in the dithered case. In addition to this, if we decided to include results for fainter magnitudes by going deeper, the lack of uniformity of the undithered field would enhance the impact of the observing conditions over the clustering signal. We can also see that, in the $\ell$ range considered in our analysis, the presence of bright stars is the dominant systematic effect. In the close neighborhood to bright stars our ability to detect faint sources diminishes. These faint sources are blended in the core or the tails of the brighter objects resulting in a lower mean number of detected sources, as shown in \figref{bright_object_masking}. The correction by weighting by the fraction of area covered in each pixel has a considerable impact at small-scales, being larger than the statistical uncertainty in this regime. Note that even though we restrict our analysis to pixels that have a relative area loss of less than 25\% the effect is very significant (more than 4 times larger than the statistical uncertainty at the scales considered). This showcases again the importance of considering the impact of blending in the small-scale regime for LSST and this issue should be carefully studied in future Data Challenges. The correction due to the presence of bright stars is comparable in both simulations. We expect that for future versions of the LSST Science Pipelines this effect will be smaller due to improvements in the deblending and measurement algorithms.

\begin{figure}
\centering
\includegraphics[width=0.45\textwidth]{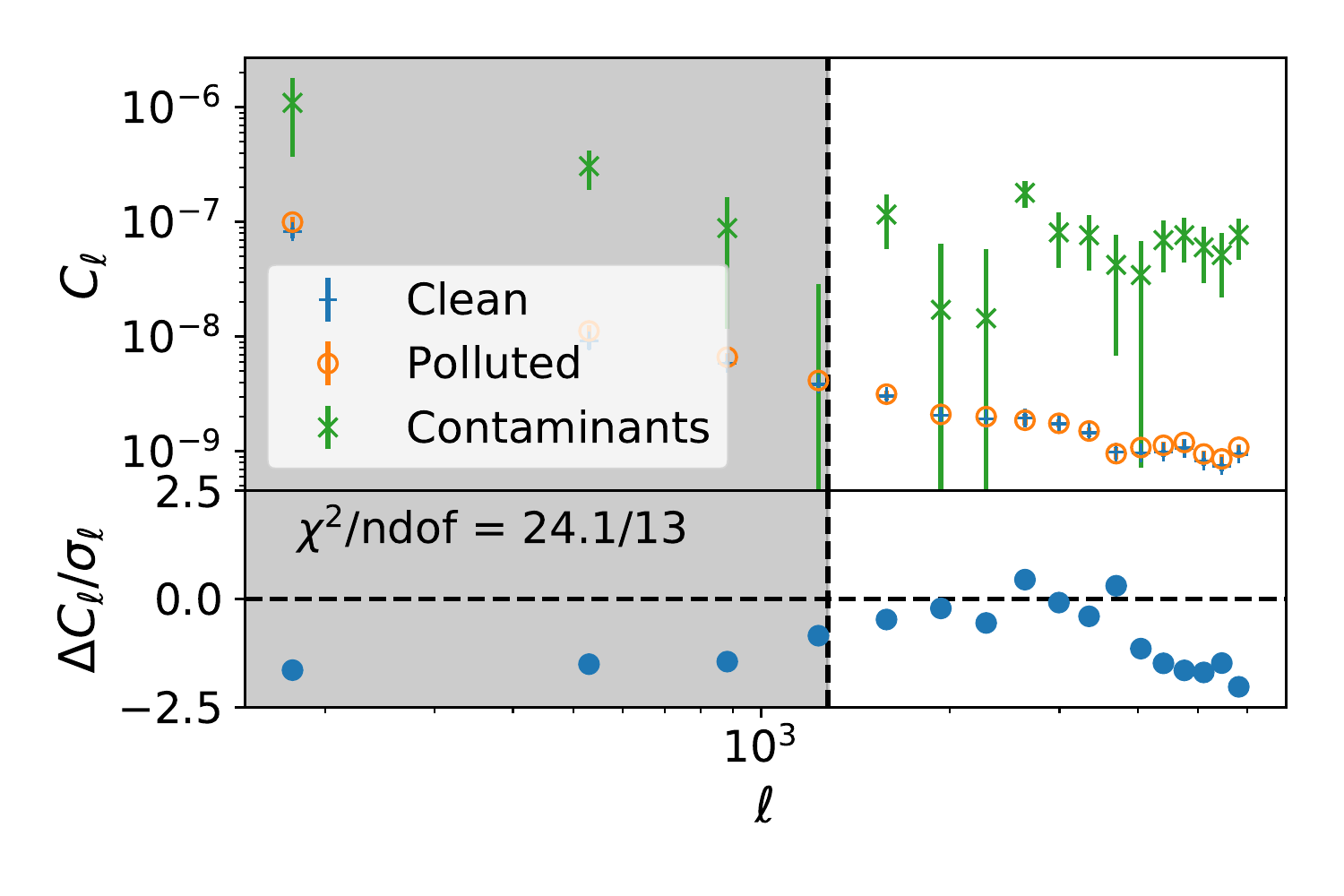}
\caption{Top: Power spectra of the clustering sample (blue $+$), the polluted sample (orange open circles), and the contaminants by themselves (green $\times$). Error bars are computed using a jackknife procedure. Bottom: Difference between the power spectra, $\Delta C_{\ell} = C_{\ell, \rm{final}} - C_{\ell, \rm{raw}}$ relative to the statistical uncertainty in the power spectra, $\sigma_{\ell}$. The vertical dashed line represents the minimum multipole used to compute the different $\ell = 1/f_{sky}$. The horizontal dashed line is drawn at zero to help visualize the difference between the power spectra. The shadowed area corresponding to $\ell < 1/f_{sky}$ is not used for our analysis.}
\label{fig:comparison_clean}
\end{figure}

Throughout this work we have used a well-behaved clean sample (our clustering sample) to demonstrate that we can successfully perform clustering measurements. However, one may think that the requirements for our sample are too restrictive. Some of the requirements described in \secref{catalogs} are indeed beyond what clustering analyses need and, as we mentioned, are driven by other science cases that will be studied with data from LSST. If artifacts are not carefully removed from the galaxy sample, we obtain significantly different clustering results. The selection cuts performed in \secref{data_selection} were motivated by the reduction of the fraction of measured sources that could not be matched to the input catalog, on top of the usual selection of a flux limited sample to ensure good uniformity across the footprint. We analyze now the power spectrum of galaxies that fulfill only the two following conditions: \texttt{r\_mag\_CModel} $\leq 25.5$ and  \texttt{base\_ClassificationExtendedness\_value} = 1. Essentially, we are adding unmatched objects to our clustering sample. We will call this sample the polluted sample. It contains 1.5\% more objects than the clustering sample. We proceed as in the previous section and compute the power spectrum deprojecting systematics and taking into account the bright object mask. We obtain the results shown in~\figref{comparison_clean}, where we can see that, although the difference in the number of objects is very small, the difference in the measured power spectra is at the $\sim 2\sigma$ level for $\ell > 3,000$. In fact, we compute $\chi^{2} = \Delta C_{\ell} C^{-1}_{\ell \ell^{\prime}} \Delta C_{\ell^{\prime}}$, with $\Delta C_{\ell} =  C_{\ell, \rm{final}} - C_{\ell, \rm{raw}}$ for $\ell, \ell^{\prime} > 1/f_{sky}$, and obtain 24.1 for 13 degrees of freedom, which means that the probability of both measurements being statistically compatible is $\sim 3\%$. Note that if we rescale the covariance to the full expected area of LSST useful for large-scale-structure studies, $\sim 14,300$ deg$^{2}$~\citep{2018arXiv180901669T}; the $\chi^{2}$/ndof becomes $\sim 9,800/13$. This means that these power spectra would be statistically incompatible, given the statistical power of LSST. In addition, we checked that these differences are not due to the change in the overall redshift distribution, $N(z)$, introduced by the newly added objects with \texttt{r\_mag\_CModel} $< 20$. These objects do change the overall $N(z)$ at the $\sim 1\%$ level. However, using the theoretical prediction given the input simulation parameters, we find that these changes lead to a $\Delta C_{\ell}/\sigma_{\ell}  < 1\%$ which is much smaller than the effect seen here. This, together with the different behavior in the power spectrum of the ``contaminants" shown in~\figref{comparison_clean}, demonstrates that the objects that we rejected by our sample selection were, indeed, artifacts that can lead to biased clustering results at small scales such as those under study in this work.

\section{Conclusions}
\label{sec:conclusions}

End-to-end simulations are powerful tools for testing the overall performance of current and future cosmological experiments like LSST. They allow us to validate and improve various aspects of the data processing and analysis, as well as to model and improve our control of systematic uncertainties. The access to ground truth allows us to test certain aspects of the processing and analysis pipelines that would be otherwise very challenging to test with real data (e.g., impact of undetected sources in the fluxes of detected overlapping sources). 

In this paper, we present an end-to-end simulated imaging dataset that resembles single-band, full-depth (10-year) LSST data (for the wide-fast-deep survey), for the first data challenge, DC1, in the LSST DESC. This dataset was generated by synthesizing sources from cosmological N-body simulations in individual sensor-visit images with different observing conditions. Two separate runs of this dataset were generated with different dither strategies, the dithered run and the undithered run. The images from both runs were processed with the LSST Science Pipelines. We find $\sim10$\% more objects in the dithered simulations due to their $\sim0.1$ magnitude deeper median depth. We perform several quality assurance tests on the resulting data products. Both datasets pass the tests performed.

We study different ways to relate the output catalogs to the inputs: The first method uses information about positions only, and the second involves both positions and magnitudes. For clustering analyses, adding information about magnitudes results in a lower incidence of spurious matches and is sufficient for DC1. However, neither of these methods provides a noiseless match between inputs and outputs.

The usage of matching strategies helps us define a galaxy sample suitable for clustering analyses. After cleaning the catalog, we find a small fraction ($\approx 3.6\%$) of artifacts, i.e., objects with no counterpart in the input. We anticipate that additional information from multi-band coverage and photometric redshifts, will help us to further refine the selection.

We perform a two-point clustering analysis of the simulated data. The results of this analysis indicate that the simulated foregrounds have a low impact, smaller than the statistical uncertainty, in both datasets. This is probably due to the simplicity of our foregrounds, and more complexity will be added in future Data Challenges. We do not find statistically significant differences in the impact of systematics between the dithered and undithered datasets given the area of DC1. We also see that for $\ell > 1150$ the presence of bright objects has a larger impact on the power spectra, $\approx 200-600\%$ of the statistical uncertainty, highlighting the impact of masking and blending with bright stars in LSST for small-scale analyses. We also demonstrate that the careful selection performed is necessary given that the presence of even a small fraction of artifacts in the sample biases the clustering results significantly.

Finally, we have been able to perform an end-to-end test of our processing and analysis pipelines. The methodology presented in this work will serve as the basis for future DESC Data Challenges, where we will aim to perform multi-band studies in a larger area, use complementary image generation strategies (PhoSim), and increase the complexity of the foregrounds included.

\appendix



\section{LSST SRD requirements}
\label{app:lsst_srd}
In this section we summarize the different requirements that we test for from the LSST Science Requirements Document version 11 that can be found at \url{https://docushare.lsst.org/docushare/dsweb/Get/LPM-17}. These requirements are driven by different science cases of LSST (for example, astrometric requirements are driven by transient science even though accurate astrometry is also important for weak lensing and galaxy clustering although not at the same level of accuracy).

\begin{table*}
\begin{tabular}{|c|c|c|c|c|c|}
\hline
KPM/Requirement & Pass/Fail Criterion & DC1 test result & Section & Figure\\
\hline
AA1 (milliarcsec) & 100 & 20 & \ref{sssec:astrometry} & \ref{fig:AA1} \\
AM1 (milliarcsec) & 20 & 8 &  & \\
AF1 (\%) & 20 & 13 &  & \\
AM2 (milliarcsec) & 20 & 4  &  & \\
AF2 (\%) & 20 & 9  &  & \\
AM3 (milliarcsec) & 30 & 7  &  & \\
AF3 (\%) & 20 & 2  &  & \\
\hline
PA1 (millimag) & 8 & 6   & \ref{sssec:photometry} & \ref{fig:validate_drp_PA1}\\
PF1 (\%) & 20 & 17   &  & \ref{fig:validate_drp_PA1}\\
\hline
PA3 (millimag) & 15 & 0.06   & \ref{sssec:zeropoints} & \ref{fig:PA34}\\
PF2 (\%) & 20 & 0  &  & \ref{fig:PA34}\\
PA6 (millimag) & 20 & 17  &  & \ref{fig:PA34}\\
\hline
D1 (mag) & 24.3 & 24.3 & \ref{sssec:depth} & \ref{fig:DF1_checks}\\
Z1 (mag) & 24.0 & 24.1 & & \ref{fig:DF1_checks}\\
DB1 (mag/r-band) & 24.3 & 24.3 &  & \ref{fig:DF1_checks}\\
Z2 (mag) & 0.4 & 0.1 &  & \\
\hline
SE1 & 0.04 & 0.001 & \ref{sssec:psf} & \ref{fig:SE1_DC1}\\
SE2 & 0.1 & 0.003 & & \ref{fig:SE1_DC1}\\
SR1 (arcsec) & 0.80 & 0.64 & & \\
SR2 (arcsec) & 1.31 & 1.01 & & \\
SR3 (arcsec) & 1.81 & 1.79 & &\\
TE1 & $3 \times 10^{-5}$ & $3\times 10^{-6}$ & & \ref{fig:TEx}\\
TE2 & $3 \times 10^{-7}$ & $9\times 10^{-8}$ & & \ref{fig:TEx}\\
\hline
WL4-Y10 (\%) & 0.1 & 0.05 & & \ref{fig:WL4-Y10}\\
\end{tabular}
\caption{Summary of Key Performance Metrics (KPMs) and DESC-SRD requirements that we check to test the quality of our end-to-end simulation pipeline. Pass/Fail criterion contains the threshold pass/fail values for each KPM/Requirement and DC1 test result shows the measured value for a given KPM/requirement. We are only able to test one criterion of the DESC-SRD,  which  is the uncertainty in the trace of the second order moments of the PSF for 10-year depth images (WL4-Y10)} 
\label{tab:kpm_table}
\end{table*}


\subsection{Astrometric requirements}
\begin{itemize}
\item AA1: Astrometric accuracy check. Minimum absolute astrometric accuracy. We compute it as the median of the difference between input and measured centroid positions. The maximum median value is 100 mas.
\item AMx: Astrometric repeatability check. Maximum RMS of the separation between pairs of stars separated by 5 arcmin (AM1: 20 mas), 20 arcmin (AM2: 20 mas), and 200 arcmin (AM3: 30 mas). 
\item AFx: Astrometric repeatability check. Maximum outlier fraction that deviate more than 40 mas (50 mas for AF3) for the separation between pairs of stars separated by 5 arcmin (AF1: 20 \%), 20 arcmin (AF2: 20\%), and 200 arcmin (AF3: 20\%).
\end{itemize} 

\subsection{Photometric requirements}
\begin{itemize}
\item PA1: Photometric repeatability check. Maximum RMS/IQR of the magnitude distribution of objects between different visits. The maximum value allowed is 8 millimags.
\item PF1: Photometric repeatability check. Maximum outlier fraction that deviate more than 15 millimag (PA2) from the mean measured magnitude. The maximum outlier fraction allowed is 20\%. 
\item PA6: Photometric accuracy check. Minimum absolute photometric accuracy. We compute this as the median difference between the input and measured fluxes for stars. The maximum allowed is 20 millimag.
\end{itemize}

\subsection{Zeropoint uniformity requirements}
\begin{itemize}
\item PA3: Zeropoint error uniformity. Maximum allowed for the RMS of the photometric zeropoint error. The maximum allowed is 15 millimags.
\item PF2: Zeropoint error uniformity. Maximum outlier fraction that deviate more than 15 millimag (PA4) in the zeropoint error distribution. The maximum allowed is 10\%
\end{itemize}

\subsection{Depth requirements}
\begin{itemize}
\item D1: Minimum depth check. Minimum value for the median of the 5$\sigma$ $r$-band depth for single visits with seeing 0.7 arcseconds, airmass 1.0 and 30 seconds exposure time. The minimum allowed is 24.3.
\item Z1: Minimum depth check. Minimum value for the 20-th percentile (DF1) of the 5-$\sigma$ $r$-band depth distribution for single-visits with seeing 0.7 arcseconds, airmass 1.0, skybrightness fainter than 21 in $r$-band, and 30 seconds exposure time.
\item DB1: Minimum depth check. DB1 is effectively the same requirement as D1, generalized to other bands (in the case of DC1 it is exactly the same as D1).
\item Z2: Minimum depth check. Maximum variation within the field of view for the brightest 20-th percentile (DF2) of the depth in a representative single visit. The maximum variation allowed is 0.4.
\end{itemize}

\subsection{Image quality requirements}
\begin{itemize}

\item SE1: Maximum PSF ellipticity check. Maximum median value of the PSF ellipticity modulus. The maximum allowed is 0.05.
\item SE2: Maximum PSF ellipticity check. Maximum value for the 90th percentile (EF1) of the PSF ellipticity modulus. The maximum allowed is 0.1.
\item SRx: Image quality test. Minimum radii to encircle at least 80\% (SR1), 95\% (SR2), and 99\% (SR3) of the flux for a fiducial delivered seeing of 0.69 arcseconds. The values are 0.80, 1.31, and 1.81 arcseconds for SR1, SR2, and SR3 respectively.
\item TE1: PSF ellipticity residual correlation check. Maximum value for the median PSF ellipticity correlations $E_{1}, E_{2}, E_{3}$ defined in equations (8-10) for $\theta \leq 1$ arcmin. The maximum allowed is $3 \times 10^{-5}$.
\item TE2: PSF ellipticity residual correlation check. Same as TE1 but for $\theta \geq 5$ arcmin. The maximum allowed is $2 \times 10^{-7}$. 
\end{itemize}
\subsection*{Acknowledgments}
We acknowledge our anonymous referee for their comments that greatly improved the quality of this work. This paper has undergone internal review in the LSST Dark Energy Science Collaboration. The internal reviewers were Andrina Nicola, Alex Drlica-Wagner and Nacho Sevilla.
FJS thanks Amanda Pagul, David Alonso, and Patricia Larsen for comments and suggestions during different stages of the manuscript. We acknowledge the use of \texttt{Pandas, Dask, SciPy, Matplotlib, Jupyter, CCL, \texttt{NaMaster}, Healpy, and scikit-learn} as well as the LSST Science Pipelines.
The work of FJS and DK was supported by the US Department of Energy award DE-SC0009920. The work of CWW was supported by the US Department of Energy High Energy Physics grant DE-SC0010007. The work of HA \& EG was supported by the US Department of Energy via grants DE-SC0011636 and DE-SC0010008. The work of JC, RD, SD, TG, AJ, HK, PJM and BVK was supported by the U.S. Department of Energy under contract number DE-AC02-76SF00515. The work of RM was supported by the US Department of Energy Cosmic Frontier progam, grant DE-SC0010118.

\input{acknowledgments}

\input{contributions}

\subsection*{Data Availability}
The data underlying this article were produced by DESC at the National Energy Research Scientific Computing Center (NERSC) and will be shared on reasonable request to the corresponding author with permission of DESC. The total data volume for DC1 is 250 TB, where $\sim$ 34 GB correspond to 10 year full-depth coadd catalogs.
\bibliography{lsstdesc,main}

\end{document}

%% file: authors.tex

\author[J.~S\'{a}nchez \textit{et. al.}]{
\parbox{\textwidth}{
\Large
J.~S\'{a}nchez,$^{1,2}$
C.~W.~Walter,$^{3}$
H.~Awan,$^{4}$
J.~Chiang,$^{5}$
S.~F.~Daniel,$^{6}$
E.~Gawiser,$^{4}$
T.~Glanzman,$^{5}$
D.~Kirkby,$^{1}$
R.~Mandelbaum,$^{7}$
A.~Slosar,$^{8}$
W.~M.~Wood-Vasey,$^{9}$
Y.~AlSayyad,$^{10}$
C.~J.~Burke,$^{11,12}$
S.~W.~Digel,$^{5}$
M.~Jarvis,$^{13}$
T.~Johnson,$^{5}$
H.~Kelly,$^{5}$
S.~Krughoff,$^{14}$
R.~H.~Lupton,$^{10}$
P.~J.~Marshall,$^{5}$
J.~R.~Peterson,$^{11}$
P.~A.~Price,$^{10}$
G.~Sembroski,$^{11}$
B.~Van Klaveren,$^{5}$
M. P.~Wiesner,$^{15}$
and B.~Xin$^{16}$
\begin{center} The LSST Dark Energy Science Collaboration \end{center}
}
\vspace{0.4cm}
\\
\parbox{\textwidth}{
$^{1}$ Department of Physics and Astronomy, University of California, Irvine, Frederick Reines Hall, Irvine, CA, U.S.A.\\
$^{2}$ Fermi National Accelerator Laboratory, P.O. Box 500, Batavia, IL, U.S.A.\\
$^{3}$ Duke University, Department of Physics, Durham, NC, U.S.A.\\
$^{4}$ Department of Physics \& Astronomy, Rutgers, The State University of New Jersey, 136 Frelinghuysen Rd, Piscataway, NJ U.S.A.\\
$^{5}$ SLAC National Accelerator Laboratory,2575 Sand Hill Rd, Menlo Park, CA, U.S.A.\\
$^{6}$ Department of Astronomy, University of Washington, Seattle, WA, U.S.A.\\
$^{7}$ McWilliams Center for Cosmology, Department of Physics, Carnegie Mellon University, Pittsburgh, PA, U.S.A.\\
$^{8}$ Brookhaven National Laboratory, Upton, NY, U.S.A.\\
$^{9}$ Pittsburgh Particle Physics, Astrophysics, and Cosmology Center (PITT PACC). Physics and Astronomy Department, University of Pittsburgh, Pittsburgh, PA 15260, U.S.A.\\
$^{10}$ Princeton University, Princeton, NJ, U.S.A.\\
$^{11}$ Department of Physics and Astronomy, Purdue University, West Lafayette, IN, U.S.A.\\
$^{12}$ Department of Astronomy, University of Illinois at Urbana-Champaign, Urbana, IL, U.S.A.\\
$^{13}$ Department of Physics and Astronomy, University of Pennsylvania, Philadelphia, PA, U.S.A.\\
$^{14}$ LSST Project Office, Tucson, AZ, U.S.A.\\
$^{15}$ Benedictine University, 5700 College Road, Lisle, IL, U.S.A.\\
$^{16}$ Large Synoptic Survey Telescope, Tucson, AZ, U.S.A.\\
}
}

%% file: acknowledgments.tex
The DESC acknowledges ongoing support from the Institut National de Physique Nucl\'eaire et de Physique des Particules in France; the Science \& Technology Facilities Council in the United Kingdom; and the Department of Energy, the National Science Foundation, and the LSST Corporation in the United States.  DESC uses resources of the IN2P3 Computing Center (CC-IN2P3--Lyon/Villeurbanne - France) funded by the Centre National de la Recherche Scientifique; the National Energy Research Scientific Computing Center, a DOE Office of Science User Facility supported by the Office of Science of the U.S.\ Department of Energy under Contract No.\ DE-AC02-05CH11231; STFC DiRAC HPC Facilities, funded by UK BIS National E-infrastructure capital grants; and the UK particle physics grid, supported by the GridPP Collaboration.  This work was performed in part under DOE Contract DE-AC02-76SF00515.
This manuscript has been authored by Fermi Research Alliance, LLC under Contract No. DE-AC02-07CH11359 with the U.S. Department of Energy, Office of Science, Office of High Energy Physics.

%% file: contributions.tex
Author contributions are listed below: \\
J.~S\'{a}nchez: led study. \\
C.~W.Walter: DC1 production planning and oversight,  led imSim development and image generation. \\
A.~Slosar: Participated in analysis and preliminary tests. \\
D.~Kirkby: Participated and advised in analysis. \\
J.~Chiang: One of the main developers of imSim, participated in image generation and data distribution. \\
T.~Glanzman: Generated artificial images. \\
S.~F.~Daniel: Developed imSim and CatSim. \\
H.~Awan: Participated in analysis and designed dithering strategy. \\
E.~Gawiser: Participated in analysis and designed dithering strategy. \\
W.~M.~Wood-Vasey: Ran validation software and wrote part of the document. \\
Y.~AlSayyad: LSST Science Pipelines \\
C.~J.~Burke: Contributions to PhoSim \\
S.~W.~Digel: Workflow and validation \\
M.~Jarvis: Contributions to imSim and validation. \\
T.~Johnson: Workflow, data generation and processing. \\
H.~Kelly: DMstack librarian at NERSC. \\
S.~Krughoff: Contributions to LSST Science Pipelines and DC1 workflow. \\
R.~H.~Lupton: LSST Science Pipelines, overall QA. \\
R.~Mandelbaum: Provided feedback on DC1 design, analysis, and paper draft. \\
P.~J.~Marshall: Led the ``Twinkles'' DC1 pathfinder project. \\
J.~R.~Peterson: PhoSim Development for DC1. \\
P.~A.~Price: LSST Science Pipelines. \\
G.~Sembroski: PhoSim Development for DC1. \\
B.~Van Klaveren: DC1 workflow. \\
M. P.~Wiesner: Completed a study on astrometry in PhoSim making it possible to improve astrometry in PhoSim for DC1. \\
B.~Xin: Contributions to imSim and PhoSim. \\